\documentclass[twocolumn]{aastex631}
\shorttitle{Brown dwarfs in $\sigma$ Orionis}
\shortauthors{Damian et al.}
\graphicspath{{./}{figures/}}

\begin{document}

\title{A novel survey for young substellar objects with the W-band filter VI: Spectroscopic census of sub-stellar members and the IMF of $\sigma$ Orionis cluster}

\correspondingauthor{Belinda Damian}
\email{belinda.damian@res.christuniversity.in}

\author[0000-0002-2234-4678]{Belinda Damian}
\affiliation{Department of Physics and Electronics, CHRIST (Deemed to be University), Hosur Road, Bengaluru 560029, India}

\author[0000-0003-4908-4404]{Jessy Jose}
\affiliation{Indian Institute of Science Education and Research (IISER) Tirupati, Rami Reddy Nagar, Karakambadi Road, Mangalam (P.O.), Tirupati 517507, India}
\email{jessyvjose1@gmail.com}

\author[0000-0003-4614-7035]{Beth Biller}
\affiliation{SUPA, Institute for Astronomy, University of Edinburgh, Blackford Hill, Edinburgh EH9 3HJ, UK}
\affiliation{Centre for Exoplanet Science, University of Edinburgh, Edinburgh, UK}

\author[0000-0002-7154-6065]{Gregory J. Herczeg}
\affiliation{Kavli Institute for Astronomy and Astrophysics, Peking University, Yiheyuan 5, Haidian Qu, 100871 Beijing, People’s Republic of China}
\affiliation{Department of Astronomy, Peking University, Yiheyuan 5, Haidian Qu, 100871 Beijing, People’s Republic of China}

\author[0000-0003-0475-9375]{Loïc Albert}
\affiliation{Institut de Recherche sur les Exoplanètes (iREx), Université de Montréal, Département de Physique, C.P. 6128 Succ. Centre-ville, Montréal, QC H3C 3J7, Canada}

\author[0000-0003-0580-7244]{Katelyn Allers}
\affiliation{Department of Physics and Astronomy, Bucknell University, Lewisburg, PA 17837, USA}

\author[0000-0002-3726-4881]{Zhoujian Zhang}
\affiliation{Department of Astronomy \& Astrophysics, University of California, Santa Cruz, 1156 High St, Santa Cruz, CA 95064, USA}

\author[0000-0003-2232-7664]{Michael C. Liu}
\affiliation{Institute for Astronomy, University of Hawai’i, 2680 Woodlawn Drive, Honolulu HI 96822, USA}

\author[0000-0001-6951-469X]{Sophie Dubber}
\affiliation{SUPA, Institute for Astronomy, University of Edinburgh, Blackford Hill, Edinburgh EH9 3HJ, UK}
\affiliation{Centre for Exoplanet Science, University of Edinburgh, Edinburgh, UK}

\author[0000-0001-6630-1971]{KT Paul}
\affiliation{Department of Physics and Electronics, CHRIST (Deemed to be University), Hosur Road, Bengaluru 560029, India}

\author[0000-0003-0262-272X]{Wen-Ping Chen}
\affiliation{Institute of Astronomy, National Central University, 300 Zhongda Road, Zhongli, Taoyuan 32001, Taiwan}
\affiliation{Department of Physics, National Central University, 300 Zhongda Road, Zhongli, Taoyuan 32001, Taiwan}

\author[0000-0003-1618-2921]{Bhavana Lalchand}
\affiliation{Institute of Astronomy, National Central University, 300 Zhongda Road, Zhongli, Taoyuan 32001, Taiwan}

\author[0000-0003-1634-3158]{Tanvi Sharma}
\affiliation{Institute of Astronomy, National Central University, 300 Zhongda Road, Zhongli, Taoyuan 32001, Taiwan}

\author{Yumiko Oasa}
\affiliation{Faculty of Education / Graduate School of Science and Engineering, Saitama University, 255 Shimo-Okubo, Sakura-ku, Saitama 338-8570, Japan}



\begin{abstract}

Low-mass stars and sub-stellar objects are essential in tracing the initial mass function (IMF). We study the nearby young $\sigma$ Orionis cluster (d$\sim$408 pc; age$\sim$1.8 Myr) using deep NIR photometric data in J, W and H-bands from WIRCam on the Canada-France-Hawaii Telescope. We use the water absorption feature to photometrically select the brown dwarfs and confirm their nature spectroscopically with the IRTF-SpeX. Additionally we select candidate low-mass stars for spectroscopy and analyze their membership and that of literature sources using astrometry from Gaia DR3. We obtain the near-IR spectra for 28 very low-mass stars and brown dwarfs and estimate their spectral type between M3-M8.5 (mass ranging between 0.3-0.01 M$_{\odot}$). Apart from these, we also identify 5 new planetary mass candidates which require further spectroscopic confirmation of youth. We compile the comprehensive catalog of 170 spectroscopically confirmed members in the central region of the cluster, for a wide mass range of $\sim$19-0.004 M$_{\odot}$. We estimate the star/BD ratio to be $\sim$4, within the range reported for other nearby star forming regions. With the updated catalog of members we trace the IMF down to 4 M$_\mathrm{Jup}$ and we find that a two-segment power-law fits the sub-stellar IMF better than the log-normal distribution.

\end{abstract}



\section{Introduction}
\label{sec:introduction}

One of the important open questions in stellar astrophysics is the form of the initial mass function (IMF) of star forming regions across diverse environmental conditions. The IMF is generally defined as the distribution of stellar mass of a star forming event at birth. A complete understanding of the IMF is required to trace the origin of star formation and its subsequent evolution and life history. \citet{salpeter1955} proposed a power-law form for the high-mass end of the IMF as dN/dM $\propto$ m$^{-\alpha}$ where the slope $\alpha$=2.35.  Following his pioneering work various other functional forms (like log-normal, multiple-power-law and tapered power-law) have been formulated to best represent the IMF (\citealt{chabrier2003,kroupa2001,demarchi2005}). The IMF appears to be universal at stellar masses (e.g. \citealt{damian2021}), however the universality of the IMF is still ambiguous towards the sub-stellar and mainly the planetary-mass regime(see review by \citealt{luhman2012}). The challenges in tracing the IMF in the substellar and planetary masses firstly lies in obtaining deep photometry in appropriate wavelengths that can identify brown dwarfs and distinguish them from field contaminants. Secondly confirming the nature of these objects with spectroscopy is crucial and additionally all of this requires a statistically robust sample of cool objects, which can essentially constrain the low mass IMF. 

Young star forming regions are ideal to study the IMF, since the dynamical processes have not significantly altered their mass distribution. Hence their present-day mass function can be taken as a proxy to the IMF. Quantifying the form of the IMF across the entire mass regime requires a complete census of the star forming region down to the substellar regime which includes brown dwarfs, planets and isolated free floating planetary mass objects (see e.g. \citealt{luhman2012,caballero2018}). Brown dwarfs form a bridge between low-mass stars and planets. These sub-stellar objects do not have a continuing nuclear energy source, hence their evolution can be simplistically pictured as a process of cooling, with decreasing luminosity as age increases \citep{chabrier2000}. Therefore young star forming regions are ideal test beds to observe these objects which are brightest during their early stages of evolution \citep{burrows1997}.

We have surveyed the $\sigma$ Orionis cluster in the solar neighbourhood, in search of low-mass stars and brown dwarfs. The $\sigma$ Orionis cluster is a young ($\sim$ 2-4 Myr, \citealt{osorio2002, sherry2008}) star forming region located near the Horsehead Nebula in the Ori OB1b association belonging to the Orion complex. This region was first described by \citet{garrison1967} and later several photometric and spectroscopic studies have been carried out (\citealt{bejar1999,bejar2001,osorio2002,kenyon2005,lodieu2009,pena2012,hernandez2014,osorio2017,caballero2019}). A  kinematic study by \citet{froebrich2021} using the photometric and astrometric data from Gaia EDR3 showed that the cluster has a  main population of young stellar objects at $\sim$403 pc. The cluster is affected by almost negligible extinction A$_\mathrm{V}$ $<$ 1 mag (\citealt{lee1968,bejar2004}), which enables the identification of the coolest objects in the region. The cluster is proposed to consist of a dense core of radius $\sim$20$\arcmin$ and a relatively sparse halo extending further to about 30$\arcmin$ \citep{caballero2008}. \citet{bejar2011} studied the spatial distribution of brown dwarfs in a  large area (1.12 deg$^2$) around the $\sigma$ Ori massive star at the center of the cluster and reported that there is no clear mass segregation of the sub-stellar objects.  

Canonical methods of sub-stellar studies are based on the initial photometric identification, wherein the candidate sources are selected based on their location in the colour-magnitude diagrams (CMD) followed by spectroscopic observation to confirm their nature \citep{pena2012}. The approach we have followed to hunt for the low-mass objects in the $\sigma$ Orionis cluster combines the J and H broad band photometry along with the data from a custom medium water band filter (6\%), called the W-band centered at 1.45 $\mu$m. The measured colors distinguish the late type objects with strong water absorption features in their spectra due to their cool atmospheres from the reddened background stars that lack water absorption in their spectra. This technique efficiently selects the cool brown dwarfs and planetary-mass members in various young star forming regions in the solar neighbourhood such as Serpens-south and Serpens-core (\citealt{jose2020,dubber2021,dubber2023}), IC 348 and Barnard 5 \citep{lalchand2022}. However, field dwarfs of late M and L type can also be selected by this method, so follow-up spectroscopy is required to confirm the youth of the candidates. 

The paper is structured as follows. In section~\ref{sec:photo_data} we present the near-infrared (NIR) photometric data used. In section~\ref{sec:candidate_selection} we discuss the  selection of candidate low-mass members of the cluster and in section~\ref{sec:spec_followup} we describe the follow-up spectroscopic observations of the selected candidate members. Section~\ref{sec:charaterization} details the spectral type classification of the members and the estimation of the parameters like extinction, age and mass. In section~\ref{sec:IMF} we present the IMF of the $\sigma$ Orionis cluster and finally in section~\ref{sec:discussion} we discuss our results and compare them with other young star forming regions.

\section{NIR photometry}
\label{sec:photo_data}
Photometric data for $\sigma$ Orionis cluster was obtained using the Wide-field InfraRed Camera (WIRCam) on the Canada-France-Hawaii Telescope (CFHT) \citep{puget2004}. The field of view of WIRCam is $\sim$ 21$\arcmin$ x 21$\arcmin$, with a sampling of 0.3 arcseconds/pixel.
The observations were carried out on 2016 September 11 (J-,H-,W-bands) and 2016 November 8 (W-band) as part of programs 16BF22 (PI: Bonnefoy) and 16BC17 (PI: Artigau) using a single pointing centered at RA = 84.700$^{\circ}$ and Dec = -2.568$^{\circ}$. We used a 21-point dithering pattern to fill the gaps between the four detectors of WIRCam and to accurately subtract the sky background. Along with the broad band J and H filters we have also used our custom filter centered on the water absorption feature at 1.45$\micron$ (W-band filter) with a total integration time of 630 s, 630 s, and 4095 s respectively. The data in all three filters were reduced using `I`iwi \footnote{\url{https://www.cfht.hawaii.edu/Instruments/Imaging/WIRCam/IiwiVersion2Doc.html}} and calibrated following the method detailed in \citet{jose2020}. The J and H band WIRCam photometry of sources with J$<12.5$mag are likely to be saturated. Hence for those sources we replace their J and H-band data with the 2MASS photometry \citep{cutri2003}. The completeness of photometry in all the three bands are detailed in appendix~\ref{sec:app_completeness}.

\section{Selection of low-mass candidates}
\label{sec:candidate_selection}
The main objectives of this work are to identify and spectroscopically confirm new low-mass members in the $\sigma$ Orionis cluster. We then build a comprehensive membership catalog for the cluster by collating the newly identified members with the known members from previous studies and present the IMF. The membership catalogue presented here for the central region of the cluster is limited by our survey field ($\sim$ 21$\arcmin$ x 21$\arcmin$). Sources outside this area are excluded from our analysis.  Within the CFHT survey field, we have identified a total of 170 members, covering a spectral range from the bright O9.5V star $\sigma$ Ori at the center of the cluster to the faintest L type brown dwarfs in the region. This membership catalog comprises of 132 known members with spectral classification, 9 that lack spectral classification from previous works, 28 low-mass members observed in this study (refer sections~\ref{sec:spec_followup} and ~\ref{sec:spectral_typing}) and 1 additional L3.5 source (S Ori 65) that is located at the edge of our survey field and due to low SNR is undetected by WIRCam. S Ori 65 was observed by \citet{barrado2001} using the FORS1 spectrograph at the Very Large Telescope. The object ID, coordinates, and photometry of all these sources along with their spectral type are provided in Table~\ref{tab:members_catalog}. The various steps followed to compile the members in this region are detailed in the following sections.

\subsection{Photometric selection based on Q-index}
\label{sec:Qindex_selection}
In order to identify and distinguish brown dwarfs and sub-stellar objects from reddened background sources, we follow the W-band technique as described in \citet{allers2020}, \citet{jose2020}, \citet{dubber2021}, \citet{dubber2023} and \citet{lalchand2022}. The W-band filter is a custom made near-IR medium band-pass filter centered at the 1.45$\micron$ water absorption feature and tailored to detect low-mass stars and brown dwarfs. The design and details of this filter are explained in \citet{allers2020}. The W-band technique uses the data from this filter combined with the data from the J and H broad band filters to classify brown dwarfs based on a reddening insensitive index (Q).

\begin{equation}\label{eq:1}
    Q = (J-W) + e(H-W)
\end{equation}

\noindent where J, W and H are the magnitudes from the corresponding filters and e is the ratio of extinction in each of these bands:

\begin{equation}\label{eq:2}
    e = \frac{A_{J}-A_{W}}{A_{W}-A_{H}}
\end{equation}

The value of e is based on the type of the commonly present contaminant in the W-band survey, the M0 stars. To reflect this, the value of e was adopted as 1.85 based on the synthetic photometry of a M0 standard spectrum from \citet{kirkpatrick2010} reddened by $A_{V}$=10 mag for $R_{v}$=3.1. The Q-index is scaled such that Q=0 corresponds to M0 type star which does not show prominent water absorption in its spectra. Accordingly, objects with later spectral types, due to deepening water absorption features will have increasing negative Q values. Using the J, W and H-band synthetic photometry of young objects (\citealt{muench2007,allers2013,allers2020}) and field dwarfs \citep{cushing2005}, the objects at the bottom of the main sequence or the hydrogen burning limit with spectral type M6 ($\sim$ 0.08 M$_{\sun}$) will have a Q=-0.6 i.e. later type objects will have a decreasing Q$<$-0.6 and earlier type objects will have Q$>$-0.6. The uncertainty in Q is estimated by propagating the errors in the J, W and H-bands in equation \ref{eq:1} as,

\begin{equation}\label{eq:3}
    \sigma_Q = \sqrt{\sigma_{J}^{2} + (e\sigma_H)^2 + ((1+e)\sigma_W)^2}
\end{equation}

Our CFHT WIRCam survey has photometric data in all three J, W and H-bands for a total of 6086 sources. In order to select the candidate very low-mass stars and brown dwarfs in the region we use a selection criteria of Q$<$-(0.6 + 3$\sigma_Q$). With this criteria we identify 59 candidate low-mass stars/brown dwarfs in the $\sigma$ Orionis cluster.

\begin{figure}
\centering
\includegraphics[scale=0.65]{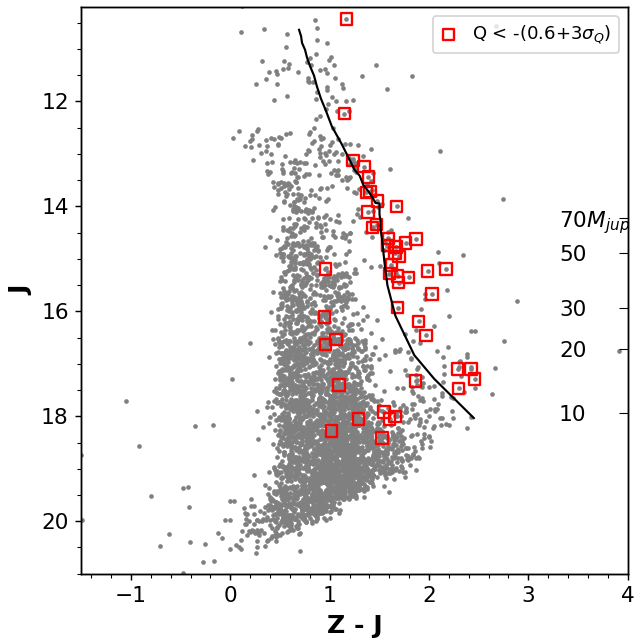}
\caption{z-J vs J CMD for the Pan-STARRS counterparts of the CFHT WIRCam data. Sources with WIRCam J$<$ 12.5 mag are replaced with 2MASS J-band photometry (refer section~\ref{sec:photo_data}). The red squares show the photometrically selected sources and the black curve shows the 2 Myr isochrone from \citet{baraffe2015} (refer section~\ref{sec:phy_param} for the estimation of age). The mass corresponding to the J-band magnitude are also shown on the right of the y-axis.} 
\label{fig:CMD_outliers}
\end{figure}

These 59 sources were further refined to remove possible field contaminants using various CMDs (a sample CMD is shown in Fig.~\ref{fig:CMD_outliers}) in order to have a robust target list for the follow-up spectroscopic observations. For this we use the optical photometry from Pan-STARRS survey (PS1) \citep{chambers2016} along with the NIR photometry from CFHT to distinguish the field sources from the low-mass stars and brown dwarf population. Forty-seven of these sources have a counterpart in the Pan-STARRS catalogue. Based on their location in the z-J vs J colour-magnitude diagram, where the majority of the sources lie along the distinct pre-main sequence branch, 11 sources lie along the field sequence towards the left of the 2 Myr isochrone (refer section~\ref{sec:phy_param}) with z-J colour difference of more than 1 mag from the isochrone. These 11 sources are considered as possible field contaminants. We also evaluate other NIR CMDs and find that apart from the 11 identified previously, one additional source is located along the field population to the left of the 2 Myr isochrone. Thus we identified 12 field contaminants based on the CMDs among the 59 Q-selected sources. 
Furthermore we checked our WIRCam survey images for possible artifacts or contaminants and found two sources located close to bright stars likely contaminating their photometry and hence were rejected based on visual inspection. Altogether 14 objects were excluded from our analysis as likely field contaminants/artifacts.

\subsection{Known members of the \texorpdfstring{$\sigma$}{} Orionis cluster}
\label{sec:known_members}

\begin{figure*}
    \centering
    \includegraphics[width=\textwidth]{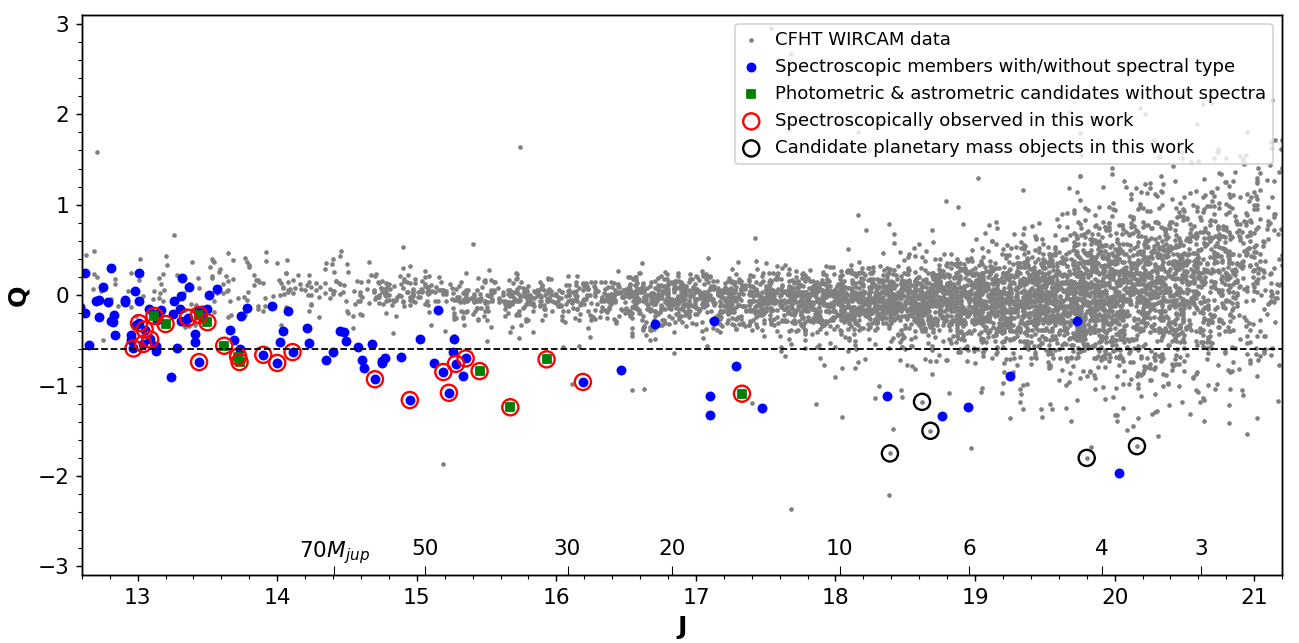}
    \caption{J vs Q plot. All the blue filled circles denote the sources which have been reported as spectroscopically known members in literature (refer section~\ref{sec:known_members}). Some of these sources do not have spectral typing in literature so have been followed up in this work. The green squares are photometrically and astrometrically identified candidates (some in this work and some identified previously) that lack spectra. Red open circles mark the sources which have been spectroscopically observed with IRTF in this work and the black open circles are newly proposed candidate planetary-mass objects. The horizontal dotted line shows the stellar/sub-stellar boundary corresponding to M6 spectral type at Q=-0.6. Any grey dot below this threshold either was rejected as field contaminant or had photometric errors too large to be selected. The mass corresponding to the J-band magnitude estimated using the mass-magnitude relation for a 2 Myr isochrone from \citet{baraffe2015} (mass$>$0.01 M$_{\sun}$) and \citet{baraffe2003} (mass$<$0.01 M$_{\sun}$) corrected for the average extinction of the cluster are also shown along the x-axis (refer section~\ref{sec:charaterization}).  }
    \label{fig:j_vs_q_knownsources}
\end{figure*}

After excluding the possible contaminants in our photometric selection, for the remaining 45 sources we checked for membership confirmation and spectral classification in the literature and found 23 of them to be spectroscopically confirmed low-mass members of the cluster and one source (WBIS\_053834.8-023930.1) to be a uncertain member. The literature membership for the sources are adopted from \citealt{bejar1999,barrado2001,bejar2001,osorio2002,barrado2003,caballero2006,caballero2007,pena2012,rigliaco2012,canty2013,hernandez2014,bozhinova2016,osorio2017,caballero2019}. Ten sources identified as members of the cluster in earlier studies based on their spectra lack estimation of their spectral class. Hence these 10 sources have been spectroscopically followed-up in this work. Among the other 11 brown dwarf candidates, 5 very faint objects have J$>$18 mag, beyond the spectroscopic sensitivity limit of a 4m class telescope. These 5 sources are listed as newly identified planetary-mass candidates at the end of Table~\ref{tab:members_catalog} and require follow-up spectroscopy. The remaining six sources along with the ten sources mentioned above and one known member (WBIS\_053855.4-024120.8) have been spectroscopically followed-up in this work (refer to section~\ref{sec:spec_followup}). All these sources are marked with different symbols in Fig.~\ref{fig:j_vs_q_knownsources}.

Additionally we compiled the known members in the region among the remaining sources in our CFHT survey. Along with WBIS\_053834.8-023930.1 mentioned above, WBIS\_053834.8-023252.3, WBIS\_053853.0-023853.7, WBIS\_053819.1-023527.9, WBIS\_053812.6-023301.5, WBIS\_053910.0-024242.4, WBIS\_053900.8-023732.0, WBIS\_053821.5-023208.4, WBIS\_053808.7-023541.3 lack reliable Gaia DR3 astrometric data (see section~\ref{sec:gaia_selection} for details on usage of Gaia DR3), hence making it difficult to ascertain their membership. Based on the literature, these sources also lack other spectral features of youth like Li I absorption line and inconsistent radial velocity (\citealt{sacco2008,hernandez2014}). We have excluded these sources from our membership catalog but suggest the reader to refer to the above works to draw conclusions based on the information provided there.  

\subsection{Member selection using Gaia DR3}
\label{sec:gaia_selection}
In this section we discuss the selection of candidate members other than those which satisfy the photometric Q-index based selection criteria (i.e. candidate members with Q$>$-0.6). To select the candidate low-mass objects for spectroscopic follow-up, we use the photometric and astrometric data from Gaia DR3 (\citealt{gaia2016,gaiacat2022}). We extract the Gaia DR3 data for our survey area ($\sim$21$\arcmin$ x 21$\arcmin$) and for those sources in our WIRCam data with counterpart in the Gaia DR3 we apply the following astrometric quality conditions: 1) renormalized unit weight error (RUWE) $<$1.4 -- sources with higher values can possibly have unreliable astrometric solutions (\citealt{kordopatis2022,stoop2022,manara2021,fabricius2021}), 2) $\sigma_\pi$/$\pi$ $<$0.1 where $\pi$ is parallax and $\sigma_\pi$ is the uncertainty in parallax (\citealt{esplin2022,penoyre2022}). We then selected for follow-up spectroscopy 11 faint sources lying along the pre-main sequence branch in the Gaia$_\mathrm{G}$ - J vs absolute Gaia$_\mathrm{G}$ CMD that did not fall within our Q-index based selection discussed in section~\ref{sec:Qindex_selection}. Since these are relatively brighter candidates they did not satisfy the Q-index based selection.  Among these 11 sources, 6 are previously spectroscopically confirmed members of the region that lacked spectral classification.

With the previously known members that satisfy the astrometric quality conditions mentioned above, we estimate the cluster parallax and proper motion. To do this, we use the weighted median of the individual parallax and proper motion of the sources respectively. The astrometric notation $\mathrm{\mu_{\alpha^*}}$ and $\mathrm{\mu_{\delta}}$
denote the proper motion in right ascension and declination where $\mathrm{\mu_{\alpha^*}}$ $\equiv$ $\mathrm{\mu_{\alpha} cos \delta}$. 

To calculate the weighted median (WM) we use the weighted.median function in the CRAN package spatstat \citep{baddeley2015} in R where 1/error$^2$ is taken as the weights \citep{kuhn2019}. The corresponding uncertainty (weighted median absolute deviation - WMAD) in parallax and proper motion is estimated using the weightedMad function in the CRAN package matrixStats. The estimated weighted median in parallax, $\mathrm{\mu_{\alpha^*}}$ and $\mathrm{\mu_{\delta}}$ are 2.45$\pm$0.12 mas, 1.23$\pm$0.91 mas yr$^{-1}$ and -0.47$\pm$0.80 mas yr$^{-1}$ respectively. From the median parallax we find the distance to the cluster to be 408$\pm$8 pc, which is consistent with other Gaia based distance estimated in \citet{monteiro2020} and \citet{froebrich2021}.

\begin{figure*}
    \centering
    \includegraphics[width=\textwidth]{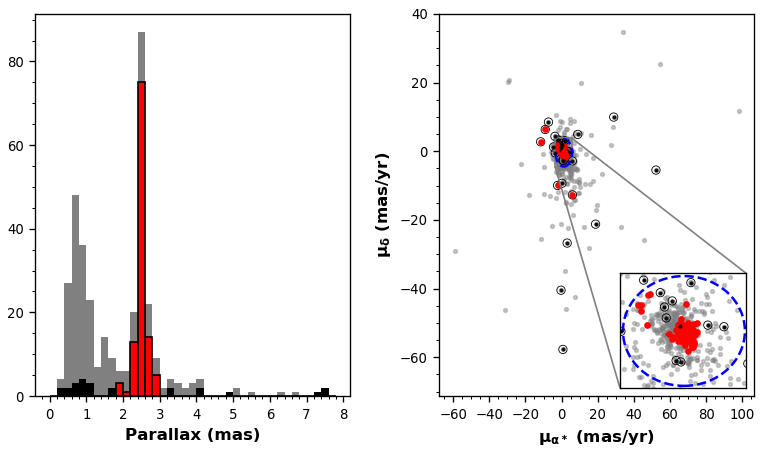}
    \caption{\textit{Left}: Histogram of parallax in units of mas. The sources in the WIRCam catalog with counterparts in the Gaia catalog and satisfying the astrometric quality conditions (RUWE$<$1.4; $\sigma_\pi$/$\pi$ $<$0.1) are shown in grey. Among the spectroscopic members of the region satisfying the above mentioned conditions and with parallax within WM$\pm$5xWMAD range are shown in red and those outside the parallax range are shown in black (refer text for details). The colour scheme is the same in both the plots. \textit{Right}: Vector point diagram with an ellipse defined by the WM and 5xWMAD values of $\mathrm{\mu_{\alpha^*}}$ and $\mathrm{\mu_{\delta}}$. The error bars represent the uncertainty in $\mathrm{\mu_{\alpha^*}}$ and $\mathrm{\mu_{\delta}}$. The black open circles mark the 28 sources excluded from the membership list.}
    \label{fig:gaia}
\end{figure*}

\begin{figure*}
    \centering
    \includegraphics[trim=70 120 70 112,clip,width=0.99\textwidth,scale=0.45]{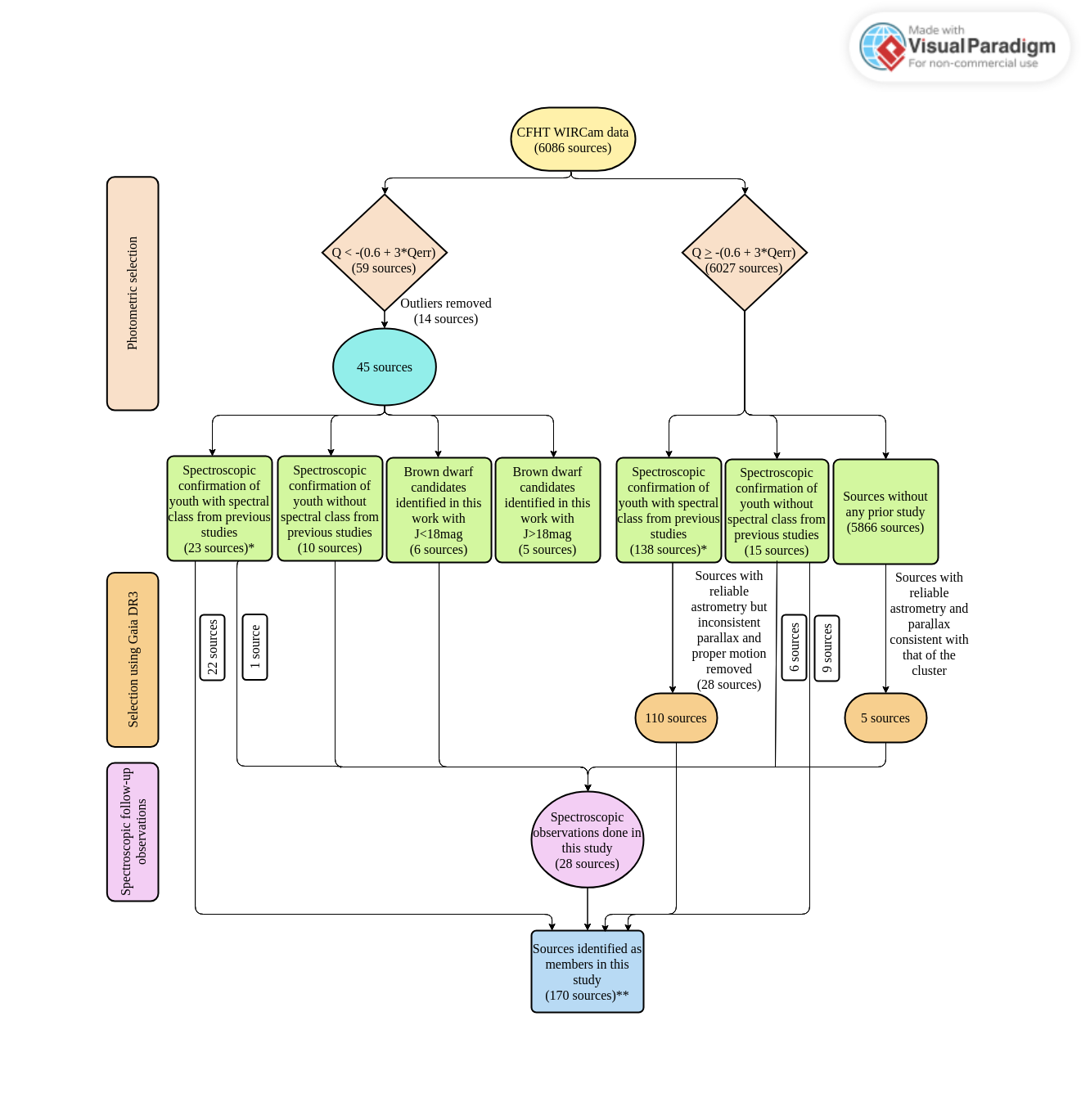}
    \caption{Flowchart depicting the various steps followed to compile the catalog of spectroscopically confirmed members of the $\sigma$ Orionis cluster. The * mark denotes that the sources mentioned as uncertain literature members are excluded and the ** indicates that this includes the low-mass member (WBIS\_053826.1-022305.0) which is located in our survey area but was undetected by WIRCam.}
    \label{fig:flowchart}
\end{figure*}

We use the estimated cluster parallax and proper motion values to further constrain the membership of the sources from previous studies through iteration. We select sources whose parallax and proper motion converges within 5$\sigma$ of their respective median values. This threshold was selected as a compromise between including as many spectroscopically identified members in the literature as possible and excluding the sources with large scatter from the median value. The left panel of Fig.~\ref{fig:gaia} shows the parallax histogram of all the sources in our WIRCam catalog with counterparts in Gaia DR3 and satisfying the astrometric quality conditions (in grey). Among the sources with which we estimated the cluster parallax, we apply a 5$\sigma$ limit on the median parallax and those which lie within the WM $\pm$ 5xWMAD range are shown in red and the remaining outliers shown in black are excluded from further study. Similarly in the vector point diagram shown in the right panel of Fig.~\ref{fig:gaia} we define an ellipse with the WM and 5xWMAD values of $\mathrm{\mu_{\alpha^*}}$ and $\mathrm{\mu_{\delta}}$ as the center and the semi-major and semi-minor axis respectively. In addition to the sources excluded as outliers in the parallax histogram we also exclude the sources lying outside the ellipse. These sources are shown as black open circles in Fig.~\ref{fig:gaia}. Therefore based on the Gaia astrometry, 28 sources with good astrometry but inconsistent parallax and proper motion have been excluded (refer Fig.~\ref{fig:flowchart}) from the membership list.

We present a flowchart in Fig.~\ref{fig:flowchart} depicting the various steps followed in compiling the membership list for the $\sigma$ Orionis cluster as explained in the previous sections.

\section{Spectroscopic follow-up observations}
\label{sec:spec_followup}
We performed spectroscopic follow-up observations of 28 low-mass candidate members in the $\sigma$ Orionis cluster (17 identified in section~\ref{sec:known_members} and 11 identified in section~\ref{sec:gaia_selection}) as marked with red open circles in Fig.~\ref{fig:j_vs_q_knownsources}. These targets were observed with the SpeX instrument on the 3.2m NASA Infrared Telescope Facility (IRTF) \citep{rayner2003} in the low-resolution prism mode (R$\sim$150) spanning a period of January 2021 to January 2022. The spectral coverage of SpeX in low-resolution mode is 0.7-2.5$\micron$. Each observation was performed using the standard ABBA nodding pattern to capture the sky and target spectra. Flat field and argon lamp observations were interspersed between target observations for the wavelength calibration. Additionally the nearby A0V standard star HD34733 was observed every hour for telluric correction and flux calibration. All the data were reduced using the IDL based spectral reduction program Spextool (version 4.1) \citep{cushing2004} following the procedure detailed in \citealt{jose2020,dubber2021}. Individual spectra from the set of target observations were extracted and wavelength calibrated using the standard star spectra. After pruning the variable effects in the flux due to bad pixels, all the target spectra were scaled to a common median flux and median combined. The standard star spectra was then used to correct for telluric absorption and flux calibration of the combined target spectra \citep{vacca2003}.

\section{Characterization of the low-mass objects}
\label{sec:charaterization}
\subsection{Spectral type and extinction from spectra}
\label{sec:spectral_typing}
\begin{figure*}
    \centering
    \includegraphics[width=\textwidth]{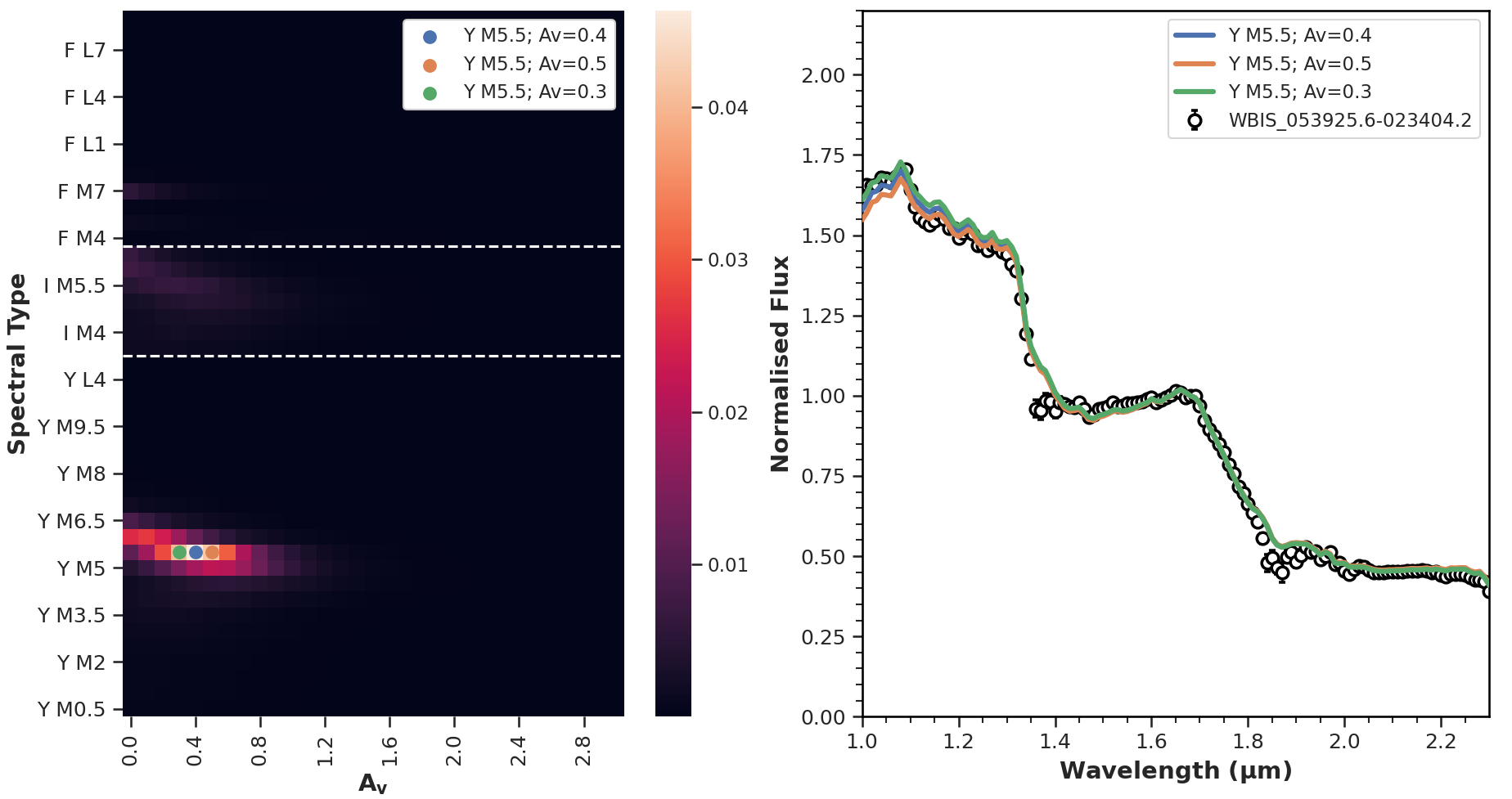}
    \caption{\textit{Left}: A$_\mathrm{V}$ vs spectral type map of WBIS\_053925.6-023404.2. The horizontal white lines demarcate the standards belonging to the three age groups. The prefix 'Y', 'I' and 'F' signify the young, intermediate age and old dwarf standards respectively. The colored markers denote the three best-fitting spectral type and A$_\mathrm{V}$ obtained from the reduced $\chi^2$ 'goodness-of-fit'. \textit{Right}: Spectra of WBIS\_053925.6-023404.2 sampled into bins of size 100\AA~ and normalised to the H-band flux at 1.68$\mu$m. The error bars show the uncertainty in the binned data. The colored lines show the three best-fitting templates. 
    }
    \label{fig:spectral_typing}
\end{figure*}

The spectral types of the candidates were estimated by comparing their spectra with the standard spectra of ultra cool dwarfs. We use three sets of standards: young standards of age $<$5 Myr; intermediate age of $\sim$10Myr (both taken from \citealt{luhman2017}); and old field dwarfs taken from the SpeX prism spectral library \citep{burgasser2014}. The spectral templates are available for young standards between M0-L7 (indicated with prefix 'Y' in this work), for intermediate age between M3.5-M6.5 (indicated with prefix 'I' in this work) and for old field dwarfs between M4-L9 (indicated with prefix 'F'). To obtain the spectral type and extinction of the candidate brown dwarfs, we find the reduced $\chi^2$ goodness-of-fit by comparing the object spectra with every standard spectra reddened by a range of A$_\mathrm{V}$ values using the \citet{fitzpatrick1999} reddening law. The extinction A$_\mathrm{V}$ varies between 0-3mag in steps of 0.1. The reduced $\chi^2$ value is calculated at each grid point along the A$_\mathrm{V}$ range for each spectral type. The three best-fitting standards and their corresponding extinction values for a sample candidate low-mass object observed with IRTF (WBIS\_053925.6-023404.2) are shown in Fig.~\ref{fig:spectral_typing}. With the estimated $\chi^2$ values, we find the normalised 1/$(\chi^2)^2$ and make an extinction vs spectral type map to represent the range of best matching templates. The right panel shows the spectra of the object normalised to the H-band flux at 1.68$\mu$m and over plotted with the three best-fitting spectral templates reddened by the corresponding A$_\mathrm{V}$. 

\begin{figure*}
    \centering
    \includegraphics[width=\textwidth]{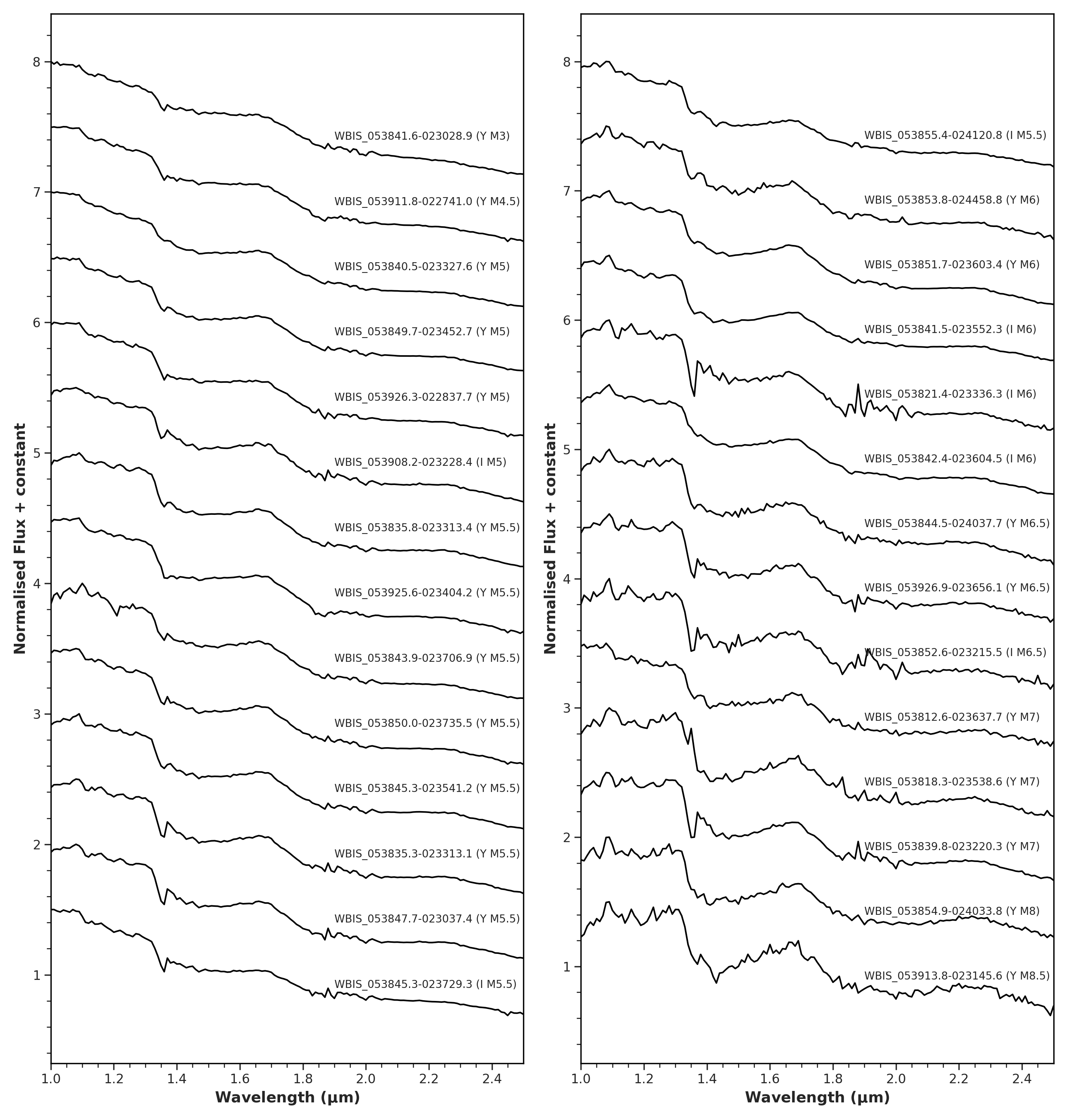}
    \caption{Normalised dereddened near IR spectra of the 28 low-mass candidate members observed with IRTF SpeX (refer sections~\ref{sec:spec_followup} and ~\ref{sec:spectral_typing} for details). Source IDs and their respective spectral types are also mentioned above each spectra. }
    \label{fig:spectra_irtf_targets}
\end{figure*}

The dereddened normalised spectra of all the 28 low-mass objects observed with IRTF are shown in Fig.~\ref{fig:spectra_irtf_targets}. All the sources have best fit templates with young or intermediate age and spectral type between M3-M8.5. The template fitting for two targets WBIS\_053855.4-024120.8 and WBIS\_053845.3-023729.3 shows that their spectra matches with the templates of both field dwarfs and intermediate age source with negligible difference between their $\chi^2$ values. Since both the targets are found to be situated along the PMS population in various CMDs and their estimated age  is $<$5 Myr (i.e. 4.5 Myr and $<$0.5 Myr respectively) (refer to section ~\ref{sec:phy_param} and Table~\ref{tab:physical_parameters}), its plausible that they are young members of the region, so we assign a dual age classification to these two sources as Y/I. The spectral type of all the 170 members -- which includes the 28 sources observed with IRTF and the 142 previously known members-- in this region are presented in Table~\ref{tab:physical_parameters}, excluding the 9 sources (with Q$>$-0.6) that lack spectral classification in literature. Our spectral type classifications have an average uncertainty of $\pm$ 1 sub-class. We also observe the 'peaky' H-band feature that is characteristic to young late type objects in the spectra of all our spectroscopically observed candidates with Q$<$-0.6. The extinction of the sources ranges between  A$_\mathrm{V}$ 0-1.8 mag with more than 90 \% of the sources having A$_\mathrm{V}$ $<$ 1 mag. The median extinction was found to be 0.4$\pm$0.5 mag.

\subsection{Spectral type based on spectral index}
\label{sec:zj_spt}
In addition to the above method where the spectral type of the 28 low-mass candidates observed with IRTF is obtained by comparing the object spectra with the standard spectra, we also use the index based spectral classification described in \citet{zhang2018}. In this section, we obtain the spectral type based on three reddening-free H$_2$O indices (H$_2$O, H$_2$OD, H$_2$O-2) where these indices are redefined based on  those modeled in \citet{allers2013}. We also estimate the youth of these objects based on gravity sensitive spectral indices (FeH$_\mathrm{z}$, VO$_\mathrm{z}$, FeH$_\mathrm{J}$, KI$_\mathrm{J}$, and H-cont), which are defined as the flux ratios of specific spectral features in the NIR spectra. The sources are assigned an age depending on the gravity index as low-gravity (VL-G), intermediate gravity (INT-G) and field gravity (FLD-G) corresponding to $<$30 Myr, $\sim$30-200 Myr and $>$200 Myr respectively. The spectral type and gravity classification for the 28 sources are given in Table~\ref{tab:members_catalog}. We see that the comparison of the object spectra with the standard template (see Section~\ref{sec:spectral_typing}) yields spectral types that are $\sim$1-2 subtypes earlier than those estimated using the spectral index. This could be due to the systematically younger spectral templates from \citet{luhman2017} than the VL-G templates adopted in \citet{zhang2018}.

\subsection{Physical parameters of members}
\label{sec:phy_param}
In order to obtain the IMF of this region, we first derive the age and mass of all the member sources. We utilise the spectral type estimated in section~\ref{sec:spectral_typing} and for the known members of the region from literature presented in Table~\ref{tab:members_catalog}. The 9 cluster members that lack spectral classification are not considered for the IMF.

For the massive early O, B, and A type objects, we incorporate for the respective spectral type the corresponding T$_\mathrm{eff}$, bolometric luminosity and mass from the extended table\footnote{\url{http://www.pas.rochester.edu/~emamajek/EEM_dwarf_UBVIJHK_colors_Teff.txt}} of \citet{pecaut2013}.

For the F to L spectral type objects we estimate the bolometric luminosity by incorporating the bolometric magnitude M$_\mathrm{bol}$ given by the below equation,

\begin{equation}
    M_\mathrm{bol}  =  m_\mathrm{J} - 5log(d) + 5 - A_\mathrm{J} + BC_\mathrm{J} 
\end{equation}

\noindent In the above equation, m$_\mathrm{J}$ corresponds to the J-band magnitude of each source, d is the distance to the cluster estimated in section~\ref{sec:gaia_selection} taken as 408 pc, and A$_\mathrm{J}$ is obtained using the mean extinction of the sources (A$_\mathrm{V}$=0.4 mag) determined in section~\ref{sec:spectral_typing} and the conversion factor from \citealt{cardelli1989}. The bolometric correction  BC$_\mathrm{J}$ is taken from \citet{pecaut2013} (for F0-M5 spectral type sources), \citet{herczeg2015} (for M5.5-M6.5 spectral type sources) and \citet{filippazzo2015} (for M7-L5 spectral type sources). The bolometric luminosity is then derived from the bolometric magnitude using the below equation where the solar bolometric magnitude M$_\mathrm{bol,\sun}$ is taken to be 4.73 mag. 

\begin{equation}
    log(L_\mathrm{bol}/L_{\sun}) = -(M_\mathrm{bol} - M_\mathrm{bol,\sun})/2.5
\end{equation}

\begin{figure}
    \centering
    \includegraphics[scale=0.65]{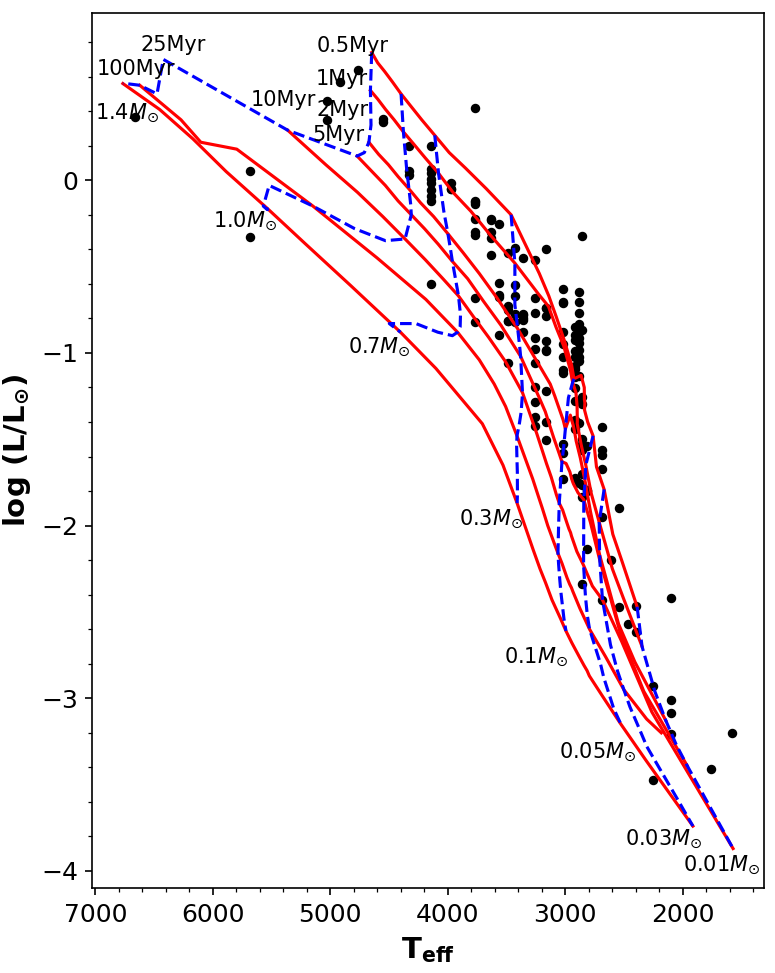}
    \caption{H-R diagram of cluster members with spectral type F to L. The blue dotted and the red continuous lines are the evolutionary tracks and isochrones from \citet{baraffe2015}. The isochrones are marked for 0.5, 1, 2, 5, 10, 25 and 100 Myr and the evolutionary tracks for 0.01, 0.03, 0.05, 0.1, 0.3, 0.7, 1 and 1.4 M$_{\sun}$.}
    \label{fig:hrd}
\end{figure}

\noindent We then obtain the T$_\mathrm{eff}$ for the F0 to M5 spectral type sources from table 6 of \citet{pecaut2013}, for the M5.5 to M6.5 spectral type sources from table 5 of \citet{herczeg2014} and for the M7 to L5 spectral type sources using the appropriate relation from \citet{filippazzo2015}. Having estimated the T$_\mathrm{eff}$ and L$_\mathrm{bol}$, we obtained the age and mass of the cluster members by charting their position in the HR-diagram using the isochrones and evolutionary tracks from \citet{baraffe2015}, as shown in Fig.~\ref{fig:hrd}. We use the isochrones for ages ranging from log(t) = 5.78 to 7.4 at an interval of 0.01 along with the isochrones for ages of 0.5, 30, 40, 50, 80, 100 Myr. For the sources lying above the 0.5 Myr isochrone, we assign the age as $<$0.5 Myr. The mass of the sources positioned above 0.5 Myr and below 100 Myr isochrones are obtained by extrapolating the evolutionary tracks. Likewise the mass of the objects above 1.4 M$_{\sun}$ track is adopted from \citet{pecaut2013} for the respective spectral type. For the very low-mass objects below 0.01 M$_{\sun}$ we incorporate the PMS stellar evolutionary model BT-Cond \citep{baraffe2003}, which caters to the lowest mass range suitable of these objects for ages 1-3 Myr. The estimated age and mass of the members along with the T$_\mathrm{eff}$ and L$_\mathrm{bol}$ are given in table~\ref{tab:physical_parameters}. 

We deduce the average age of the cluster by calculating the median of the individual cluster members with age above 0.5 Myr and mass above 0.01 M$_{\sun}$. The median age of the cluster is 1.8 Myr and the  median absolute deviation is taken as the corresponding uncertainty of 1 Myr. The estimated age of the cluster is consistent with the estimate of 2--4 Myr for a distance of 385 pc \citep{osorio2002} and 2--3 Myr for a distance of 444 pc \citep{sherry2008}.

\section{Initial Mass Function}
\label{sec:IMF}
\begin{figure}
    \centering
    \includegraphics[scale=0.6]{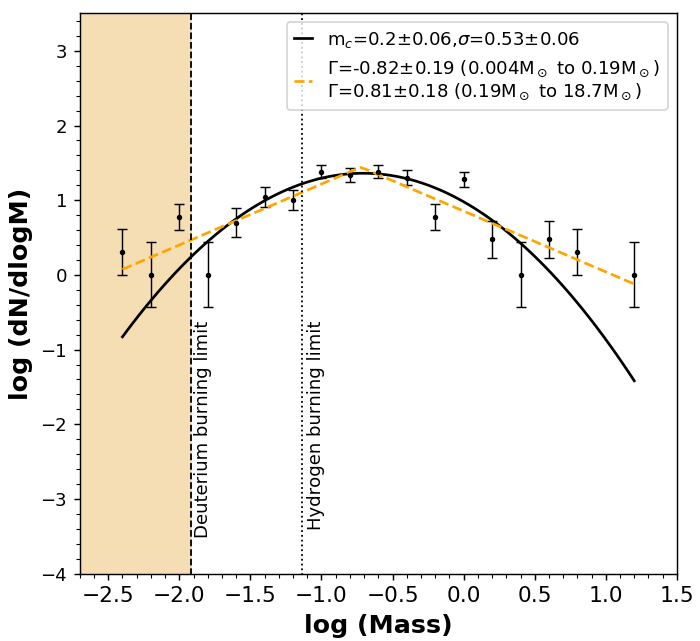}
    \caption{The IMF for the 161 members of ${\sigma}$ Ori (listed in Table ~\ref{tab:physical_parameters}) whose mass estimate was derived in section ~\ref{sec:phy_param}. The bin size for the x-axis is 0.2 dex. The error bars denote the Poisson error in each point. The continuous black curve is the log-normal fit to the IMF with the peak mass ($m_c$) = 0.20$\pm$0.06 and ${\sigma}$ = 0.53$\pm$0.06. The dashed orange lines are the segmented power-law fits to the IMF with a break-point mass at 0.19 M$_{\sun}$. The power-law fit for the low-mass end seems to better represent the IMF than the log-normal curve. The shaded region highlights the objects below the deuterium burning limit. }
    \label{fig:imf}
\end{figure}

In this section we discuss the IMF of the $\sigma$ Orionis cluster using the catalog of confirmed members compiled in this work. The mass estimated in the previous section (section~\ref{sec:phy_param}) for 161 sources (excluding the 9 sources without spectral type) are used to build the IMF across a wide spectrum of mass ranging from $\sim$19 M$_{\sun}$ to 0.004 M$_{\sun}$. The IMF shown in Fig.~\ref{fig:imf} is calculated from the number of sources in each logarithmic mass bin of size 0.2 dex, with Poisson error bars. There are various functional forms in literature that trace the IMF over different mass ranges. \citet{kroupa2001} gave the multi-segment power-law fit to the IMF of the form,

\begin{equation}
    \frac{dN}{dlogm} \propto m^{-\Gamma}
\end{equation}

\noindent which traces the distribution of the number of stars in a logarithmic mass bin of range logm + dlogm with slope ($\Gamma$). Another commonly used formalism is the  log-normal distribution of the IMF proposed by \citet{chabrier2003}  of the form,

\begin{equation}
    \frac{dN}{dlogm} \propto e^{-\frac{(logm-logm_c)^{2}}{2\sigma^{2}}}
\end{equation}

\noindent where m$_c$ is the characteristic mass (the mass at the peak of the distribution) and $\sigma$ is the standard deviation.
 Here we use both these forms to describe the IMF of the $\sigma$ Orionis cluster. The log-normal distribution to the complete mass range for all the 161 members between 18.7 M$_{\sun}$ to 0.004 M$_{\sun}$ yields a characteristic mass (or the peak mass) $m_c$=0.20$\pm$0.06 and $\sigma$=0.53$\pm$0.06. This is shown as a continuous black curve in Fig.~\ref{fig:imf}. The IMF obtained for this region is in general comparable with that of other star forming regions and the Galactic field mass function (see \citealt{chabrier2003, dario2012, suarez2019, damian2021} and references therein). We also fit a two-segment power-law fit to the IMF with a break mass at 0.19 M$_{\sun}$. The high-mass end of the IMF i.e. between 0.19-18.7 M$_{\sun}$ has a slope $\Gamma$=0.81$\pm$0.18 and the low-mass end of the IMF i.e. between 0.004-0.19 M$_{\sun}$ gives $\Gamma$=-0.82$\pm$0.19 as shown with the orange dotted lines in Fig.~\ref{fig:imf}. Our robust sample of spectroscopically confirmed members of the $\sigma$ Orionis cluster facilitates in tracing the sub-stellar IMF down to few planetary masses making it one of the deepest IMF presented for this region.

We also estimate the ratio between the stellar and brown dwarf population in our sample using the mass calculated in section~\ref{sec:phy_param}. To calculate the star/BD ratio we use the complete mass range of our catalog with 0.07 M$_{\sun}$ (hydrogen burning mass limit or the stellar/sub-stellar boundary mass at solar metallicity) as the demarcation between the two classes of objects. The lower mass limit for the brown dwarfs is taken as 0.01 M$_{\sun}$ (deuterium burning mass limit). The star/BD ratio is 4.37 i.e. $\sim$4 stars for every brown dwarf in the region. This is similar to the ratio found in other nearby young regions. \citet{luhman2016} estimated the brown dwarf to star ratio (N($\geq$M6.5)/N($<$M6.5)) for the nearby IC 348 cluster as 0.188$\mathrm{^{+0.025}_{-0.02}}$. For the Chamaeleon-I region \citet{luhman2007} computed the ratio (N(0.02-0.08 M$_{\sun}$)/N(0.08-10 M$_{\sun}$)) as 0.26$\mathrm{^{+0.06}_{-0.05}}$. \citet{andersen2008} compiled the ratio of stars to brown dwarfs in some of the well-studied young clusters like Taurus, IC 348, Mon R2, Chameleon-I, Pleiades, ONC and NGC 2024 and found the range of R (N(1-0.08 M$_{\sun}$)/N(0.08-0.03 M$_{\sun}$)) between $\sim$3-8. We note that a caveat in comparing the star/BD ratio between different regions is that the estimate depends on  the method used to derive the mass of the objects and the mass ranges considered.

\section{Discussion}
\label{sec:discussion}
\subsection{Previous studies on \texorpdfstring{$\sigma$}{} Orionis cluster}
Since the discovery of this region, with the advancement of the observing capabilities and wide-field deep surveys several studies have observed the faintest cool objects and attempted to present a complete sub-stellar IMF. Here we discuss some of the earlier studies which present the IMF of $\sigma$ Orionis cluster. \citet{caballero2007} surveyed 790 arcmin$^2$ area of the $\sigma$ Ori cluster in the I- and J-bands and with a sample of 49 brown dwarfs and planetary mass objects they found a rising mass spectrum in the mass range 0.11-0.006 M$_{\sun}$ ($\alpha$=0.6$\pm$0.2; dN/dM$\propto$M$^{-\alpha}$ where $\alpha$=$\Gamma$+1) which is slightly steeper than the slope estimated in this study. They argue that in the observed mass range they do not find any significant evidence for a mass limit to the opacity limited fragmentation and that brown dwarfs and IPMOs form via the same mechanism as low-mass stars.

\citet{lodieu2009} analysed 0.78 deg$^2$ area observed by the UKIDSS Galactic Clusters Survey and presented the IMF for the sub-stellar mass range. They derived a power-law index $\alpha$=0.5$\pm$0.2 in the mass range 0.5-0.01 M$_{\sun}$, which is in agreement with the slope for the Upper Sco region over the same mass range. A similar study by \citet{bejar2011} for a larger 1.12 deg$^2$ area found a rising slope from 0.1 M$_{\sun}$ to the deuterium burning mass limit with $\alpha$=0.7$\pm$0.3 for an age of 3 Myr. They compared their results with other young open clusters and conclude that the sub-stellar IMF is consistent.

Later a deep photometric survey by \citet{pena2012} using VISTA Orion data for a 30$\arcmin$ radius around the central massive star computed the IMF for the mass range between 20 M$_{\sun}$ to 0.006 M$_{\sun}$ where only 70\% of their sources were confirmed members. Their log-normal fit to the complete mass range produces a $m_c$=0.27$\pm$0.1 M$_{\sun}$ and variance ranging between 0.5 to 0.7. Subsequently they observe that the cluster mass spectrum is best fitted by a two-segment power-law with $\alpha$=1.7$\pm$0.2 (for M$>$0.35 M$_{\sun}$), and $\alpha$=0.6$\pm$0.2 (for M$<$0.35 M$_{\sun}$). We note that with our comprehensive catalog of all confirmed members, the mass range for the IMF closely matches with that studied by \citet{pena2012}, so a fair comparison of the two results show that both the log-normal and the two-segmented power-law fits are consistent within the uncertainties.

\subsection{Comparison with other star forming regions}
Exploring the substellar and planetary mass IMF in various star forming regions has been of interest due to its ambiguity as discussed in section~\ref{sec:introduction}. Tracing this very low mass end of the IMF in distant regions is challenging and limited by detecting these faint objects. Here we discuss some of the well studied regions within 1 kpc where the substellar IMF has been probed. In the solar neighbourhood, Upper Scorpius is one of the closest (d$\sim$145 pc; age$\sim$7-15 Myr) star forming regions. \citet{cook2017} explored the very low mass objects (VLMOs) in this region and traced the substellar IMF in the mass range 0.01-0.1 M$_{\sun}$. They found that the slope of the mass function ($\alpha \sim$ 0.4) is consistent with the Kroupa IMF.

The Orion Nebula Cluster (ONC) (d$\sim$400 pc \citep{kuhn2019}; age$\sim$1-3 Myr \citep{dario2012}) has been extensively studied to understand star formation especially its low-mass star and brown dwarf population. \citet{dario2012} conducted a series of studies on this region and they present the complete IMF between $\sim$3-0.03 M$_{\sun}$. They trace both the \citet{chabrier2003} log-normal form and the \citet{kroupa2001} power-law fits to the distribution using different evolutionary models to estimate the mass of the sources. For the \citet{baraffe1998} model and the log-normal fit they find the peak mass log $m_c$=-0.45$\pm$0.02 and $\sigma$=0.44$\pm$0.05. The two-phase power law shows a break mass at 0.29 M$_{\sun}$ with slopes -1.12$\pm$0.9 and 0.6$\pm$0.33. 

Another young region at a similar distance is the 25 Ori (d$\sim$350 pc; age$\sim$7-10 Myr \citep{suarez2019}) in the Orion OB1a association. \citet{suarez2019} studied the system IMF of this region for different spatial extent and wide mass range (0.01-13 M$_{\sun}$). Their results for the log-normal fit ($m_c$=0.31$\pm$0.06, $\sigma$=0.51$\pm$0.08) as well as the two-segment power-law fit ($\Gamma$=-0.77$\pm$0.06 for m$<$0.4 M$_{\sun}$, $\Gamma$=1.33$\pm$0.12 for m$>$0.4 M$_{\sun}$) are comparable with our estimate for $\sigma$ Orionis (see Table 6 of their paper). Beyond the solar neighbourhood within 1 kpc, NGC 2264 (d$\sim$722pc \citep{flaccomio2023}; age$\sim$0.5-5 Myr \citep{venuti2019}) is one of the well studied young clusters. Recently \citet{pearson2021} analysed the sub-stellar IMF and reported a slope of 0.43$^{+0.41}_{-0.56}$ pointing towards a universal IMF in the brown dwarf regime in comparison to other nearby clusters. 
 
A direct comparison between different studies is challenging considering the factors such as - difference in the mass ranges considered in each IMF, evolutionary models used for estimating the mass, and completeness of the sample. However we see that our estimate of the stellar/sub-stellar IMF for the $\sigma$ Ori region is fairly consistent with some of the  young star forming regions as discussed above. Based on our results, the power-law fit represents the distribution of sub-stellar objects in the low-mass end of the IMF better than the log-normal fit which underestimates the brown dwarfs and planetary mass objects in the region.

\section{Conclusions}
\label{sec:conclusion}
Brown dwarfs and sub-stellar objects in general are a unique class of sources in a young star forming region. We have implemented a novel technique to photometrically identify these cool objects in the nearby $\sigma$ Orionis cluster using a custom water band filter along with J and H broad band filters in the CFHT-WIRCam. We use the optical and NIR CMDs to remove the field contaminants and the Gaia DR3 astrometry to select candidate low-mass members of the cluster. These candidate low-mass objects were spectroscopically observed with the Spex instrument in IRTF. All the 28 candidates observed are confirmed as low-mass members or brown dwarfs with spectral type between M3 and M8.5. Our 100\% confirmation rate in photometrically identifying sub-stellar objects in the $\sigma$ Orionis cluster (all the candidates selected based on Q$<$-0.6 have spectral type $>$M5) proves that this is a robust and efficient method. 

We have compiled a catalog of 170 members in the central region of the cluster spanning a mass range of $\sim$19 M$_{\sun}$ to 4 M$_\mathrm{Jup}$ along with their spectral type and physical parameters. With our robust census of spectroscopic members we build the IMF and trace it down to the sub-stellar and planetary mass regime. This is one of the deepest IMF available for the $\sigma$ Orionis cluster using a sample of spectroscopically confirmed members. The log-normal fit to the IMF yields a  peak characteristic mass ($m_c$) 0.20$\pm$0.06 and $\sigma$=0.53$\pm$0.06 which is consistent with other studies of nearby star forming regions. The two-segment power-law fit breaks at 0.19 M$_{\sun}$ with a slope -0.82$\pm$0.19 for the low-mass end and 0.81$\pm$0.18 for the high mass end. We observe that the segmented power law fit represents the sub-stellar IMF better than the log-normal distribution. We also estimate the star to brown dwarf ratio to be $\sim$4. We conclude that our estimate of the IMF for the $\sigma$ Orionis cluster is consistent with other nearby star forming regions.

\begin{acknowledgments}
The authors thank the anonymous referee for the constructive report which has helped improve the overall quality of the paper. Based on observations obtained with WIRCam, a joint project of CFHT, Taiwan, Korea, Canada, and France, at the
Canada–France–Hawaii Telescope (CFHT) which is operated by the National Research Council (NRC) of Canada, the Institut National des Sciences de l’Univers of the Centre National de la Recherche Scientifique of France, and the University of Hawaii. Visiting Astronomer at the Infrared Telescope Facility, which is operated by the University of Hawaii under contract 80HQTR19D0030 with the National Aeronautics and Space Administration. This research has benefitted from the SpeX Prism Spectral Libraries, maintained by Adam Burgasser at \url{http://www.browndwarfs.org/spexprism}. This work has made use of data from the European Space Agency (ESA) mission
{\it Gaia} (\url{https://www.cosmos.esa.int/gaia}), processed by the {\it Gaia} Data Processing and Analysis Consortium (DPAC, \url{https://www.cosmos.esa.int/web/gaia/dpac/consortium}). Funding for the DPAC has been provided by national institutions, in particular the institutions participating in the {\it Gaia} Multilateral Agreement. This publication makes use of data products from the Two Micron All Sky Survey, which is a joint project of the University of Massachusetts and the Infrared Processing and Analysis Center/California Institute of Technology, funded by the National Aeronautics and Space Administration and the National Science Foundation.
This research has made use of the SIMBAD database, operated at CDS, Strasbourg, France and the cross-match service provided by CDS, Strasbourg. BD is thankful to the Center for Research, CHRIST (Deemed to be University), Bangalore, India. BD is also thankful to Isabelle Baraffe for providing the evolutionary models through personal communication and to Shridharan Baskaran for the valuable suggestions and discussions. 
BD and JJ  acknowledge the support rendered by the IRTF team especially Bobby Bus, Michael Connelley and Adwin Boogert for the training on observing with IRTF; Bernie Walp for the help and guidance during the observing run and Tony Denault for granting access to the IRTF Freia system to reduce the data.  GJH is supported by general grant 12173003 National Natural Science Foundation of China.
\end{acknowledgments}

%

\vspace{5mm}
\facilities{CFHT (WIRCam), IRTF (SpeX), Gaia}


\software{astropy \citealt{astropy2013,astropy2018},  
          Spextool \citep{cushing2004}
          }



\appendix

\section{Completeness of the CFHT WIRCam data}
\label{sec:app_completeness}
\begin{figure}[h!]
    \centering
    \includegraphics[scale=0.65]{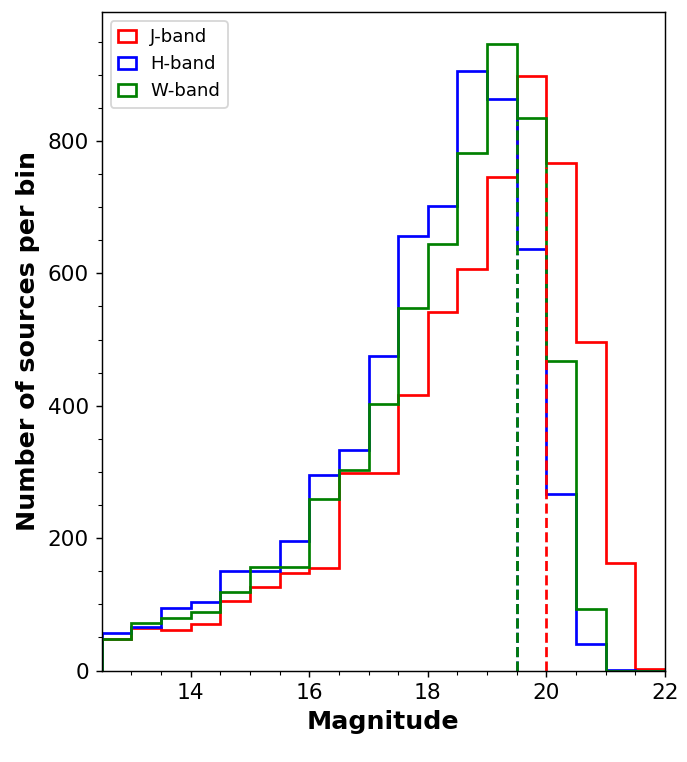}
    \caption{Completeness histograms of the WIRCam data in J-,W- and H-bands. The vertical dashed lines show the 90 \% completeness in the respective bands. }
    \label{fig:completeness}
\end{figure}

Our survey of sub-stellar objects in the $\sigma$ Orionis cluster is largely limited by the completeness of the WIRCam photometric data used. In order to evaluate the completeness of our WIRCam survey, we plot histograms of the data in the individual J, W and H bands. The turn-over point of the distribution is generally considered as a proxy for the 90 \% completeness of the data (\citealt{willis2013,jessy2017,damian2021}). In some cases when the bin trailing the turn-over point has more than 90 \% of the number of sources in the peak bin then that particular trailing bin is considered as the 90 \% completeness value. For the J, W and H bands we find the completeness to be 20 mag, 19.5 mag and 19.5 mag respectively (as shown in Fig.~\ref{fig:completeness}) which corresponds to $\sim$0.004 M$_{\sun}$ ($M_J$=11.86mag) in all the three bands considering their respective extinction, mean distance to the cluster and the 2 Myr BT-Cond evolutionary model \citep{baraffe2003} (see details in section~\ref{sec:charaterization}). The completeness of our membership catalog is also limited by the photometric selection of brown dwarfs using the Q-index (refer section~\ref{sec:Qindex_selection}). At the CFHT WIRCam J-band completeness of 20 mag, considering the median error of the sources in Q, two sources satisfy the selection criteria of Q$<$-(0.6+3$\sigma_{Q}$) among which one is a known member in literature and has been listed in our catalog.

\section{Membership catalog of sources in the \texorpdfstring{$\sigma$}{} Orionis cluster}
\label{sec:app_memcat}
\begin{longrotatetable}
\begin{deluxetable*}{cccccccccc}
    \tablecaption{Complete membership catalog of sources in the $\sigma$ Orionis cluster for a $\sim$21$\arcmin$ x 21$\arcmin$ survey area around the central massive star. }
    \label{tab:members_catalog}
    \tabletypesize{\scriptsize}
    \tablehead{
     \colhead{Object ID} & \colhead{RA} & \colhead{Dec} & \colhead{J} & \colhead{[1.45]} & \colhead{H}  & \colhead{Q} & \colhead{SIMBAD name$^a$} & \colhead{SpT} & \colhead{References}\\
     \colhead{} & \colhead{(deg)} & \colhead{(deg)} & \colhead{(mag)} & \colhead{(mag)} & \colhead{(mag)} & \colhead{} & \colhead{} & \colhead{}\\
     }
     \startdata
     \multicolumn{10}{l}{\textit{Sources with spectroscopic membership from previous works $^b$:}}\\[1ex]
     WBIS\_053844.8-023600.2$^*$	&	84.6865	&	-2.6000	&	$4.75\pm0.26$	&	-	&	$4.64\pm0.25$	&	-	&	* sig Ori	&	O9.5V	&	14	\\
    WBIS\_053847.2-023540.6$^*$	&	84.6967	&	-2.5946	&	$6.97\pm0.03$	&	-	&	$6.95\pm0.03$	&	-	&	* sig Ori E	&	B2	&	14	\\
    WBIS\_053845.6-023559.0$^*$	&	84.6902	&	-2.5997	&	$7.12\pm0.03$	&	-	&	$7.22\pm0.03$	&	-	&	* sig Ori D	&	B2	&	14	\\
    WBIS\_053836.5-023312.7$^*$	&	84.6523	&	-2.5535	&	$8.10\pm0.02$	&	-	&	$8.18\pm0.06$	&	-	&	HD 294271	&	B5	&	14	\\
    WBIS\_053901.5-023856.4$^*$	&	84.7562	&	-2.6490	&	$8.13\pm0.03$	&	-	&	$8.11\pm0.04$	&	-	&	HD 37525	&	B5	&	14	\\
    WBIS\_053834.8-023415.7$^*$	&	84.6450	&	-2.5710	&	$8.35\pm0.03$	&	-	&	$8.38\pm0.05$	&	-	&	HD 294272A	&	A1	&	14	\\
    WBIS\_053834.2-023416.0$^*$	&	84.6426	&	-2.5711	&	$8.78\pm0.02$	&	-	&	$8.82\pm0.07$	&	-	&	HD 294272B	&	B7	&	14	\\
    WBIS\_053844.1-023606.3$^*$	&	84.6839	&	-2.6018	&	$9.09\pm0.03$	&	-	&	$9.11\pm0.05$	&	-	&	* sig Ori C	&	A2	&	14	\\
    WBIS\_053838.5-023455.0$^*$	&	84.6603	&	-2.5820	&	$9.91\pm0.03$	&	-	&	$9.28\pm0.02$	&	-	&	[W96] 4771-1147	&	K2	&	14	\\
    WBIS\_053833.7-024414.3$^*$	&	84.6405	&	-2.7373	&	$10.13\pm0.03$	&	-	&	$9.28\pm0.02$	&	-	&	V* TX Ori	&	K1	&	14	\\
    WBIS\_053848.0-022714.2$^*$	&	84.7001	&	-2.4540	&	$10.16\pm0.02$	&	-	&	$9.46\pm0.03$	&	-	&	[W96] 4771-899	&	M0	&	14	\\
    WBIS\_053827.5-024332.5$^*$	&	84.6147	&	-2.7257	&	$10.18\pm0.03$	&	-	&	$10.10\pm0.03$	&	-	&	HD 294273	&	A2	&	14	\\
    WBIS\_053835.9-024351.1$^*$	&	84.6495	&	-2.7309	&	$10.45\pm0.03$	&	-	&	$9.73\pm0.02$	&	-	&	V* TY Ori	&	K0	&	14	\\
    WBIS\_053807.8-023130.7$^*$	&	84.5327	&	-2.5252	&	$10.57\pm0.03$	&	-	&	$9.93\pm0.02$	&	-	&	2MASS J05380784-0231314	&	K3	&	14	\\
    WBIS\_053853.4-023323.0$^*$	&	84.7224	&	-2.5564	&	$10.61\pm0.03$	&	-	&	$9.92\pm0.02$	&	-	&	[W96] 4771-1049	&	K3	&	14	\\
    WBIS\_053918.1-022928.5$^*$	&	84.8253	&	-2.4912	&	$10.72\pm0.03$	&	-	&	$10.27\pm0.03$	&	-	&	2MASS J05391807-0229284	&	K0	&	14	\\
    WBIS\_053932.6-023944.0$^*$	&	84.8857	&	-2.6622	&	$10.82\pm0.03$	&	-	&	$10.10\pm0.02$	&	-	&	2MASS J05393256-0239440	&	K5	&	14	\\
    WBIS\_053844.2-023233.6$^*$	&	84.6843	&	-2.5427	&	$10.88\pm0.03$	&	-	&	$10.16\pm0.02$	&	-	&	2MASS J05384424-0232336	&	K4	&	14	\\
    WBIS\_053838.2-023638.5$^*$	&	84.6593	&	-2.6107	&	$11.16\pm0.03$	&	-	&	$10.47\pm0.02$	&	-	&	[W96] rJ053838-0236	&	K5	&	14	\\
    WBIS\_053834.3-023500.1$^*$	&	84.6430	&	-2.5834	&	$11.22\pm0.03$	&	-	&	$10.57\pm0.03$	&	-	&	[BNM2013] 92.01 690	&	K5	&	14	\\
    WBIS\_053859.5-024508.0$^*$	&	84.7481	&	-2.7522	&	$11.22\pm0.03$	&	-	&	$10.98\pm0.02$	&	-	&	2MASS J05385955-0245080	&	F3	&	14	\\
    WBIS\_053835.9-023043.3$^*$	&	84.6494	&	-2.5120	&	$11.25\pm0.03$	&	-	&	$10.60\pm0.02$	&	-	&	[W96] 4771-1097	&	K4	&	14	\\
    WBIS\_053907.6-023239.1$^*$	&	84.7817	&	-2.5442	&	$11.30\pm0.03$	&	-	&	$10.57\pm0.02$	&	-	&	Haro 5-19	&	K5	&	14	\\
    WBIS\_053835.5-023151.6$^*$	&	84.6478	&	-2.5310	&	$11.30\pm0.03$	&	-	&	$10.63\pm0.02$	&	-	&	[W96] rJ053835-0231	&	K4	&	14	\\
    WBIS\_053925.2-023822.0$^*$	&	84.8550	&	-2.6394	&	$11.31\pm0.03$	&	-	&	$10.45\pm0.02$	&	-	&	V* BG Ori	&	K7	&	14	\\
    WBIS\_053844.2-024019.7$^*$	&	84.6843	&	-2.6721	&	$11.36\pm0.03$	&	-	&	$10.69\pm0.02$	&	-	&	[W96] 4771-1051	&	K5	&	14	\\
    WBIS\_053849.2-023822.3$^*$	&	84.7049	&	-2.6395	&	$11.39\pm0.03$	&	-	&	$10.66\pm0.02$	&	-	&	[W96] rJ053849-0238	&	K7	&	14	\\
    WBIS\_053918.8-023053.2$^*$	&	84.8285	&	-2.5148	&	$11.40\pm0.03$	&	-	&	$10.64\pm0.03$	&	-	&	Haro 5-21	&	K7	&	14	\\
    WBIS\_053841.3-023722.6$^*$	&	84.6720	&	-2.6229	&	$11.46\pm0.03$	&	-	&	$10.79\pm0.03$	&	-	&	UCAC2 30800097	&	K5	&	14	\\
    WBIS\_053840.3-023018.5$^*$	&	84.6678	&	-2.5051	&	$11.51\pm0.03$	&	-	&	$10.76\pm0.02$	&	-	&	Haro 5-13	&	M0	&	14	\\
    WBIS\_053831.6-023514.9$^*$	&	84.6316	&	-2.5875	&	$11.52\pm0.03$	&	-	&	$10.71\pm0.02$	&	-	&	[W96] rJ053831-0235	&	M0	&	14	\\
    WBIS\_053832.8-023539.2$^*$	&	84.6368	&	-2.5942	&	$11.54\pm0.03$	&	-	&	$10.90\pm0.02$	&	-	&	[W96] rJ053832-0235b	&	M0	&	14	\\
    WBIS\_053905.4-023230.3$^*$	&	84.7725	&	-2.5417	&	$11.55\pm0.03$	&	-	&	$10.86\pm0.02$	&	-	&	[W96] 4771-1075	&	K5	&	14	\\
    WBIS\_053911.6-023602.8$^*$	&	84.7985	&	-2.6008	&	$11.62\pm0.03$	&	-	&	$10.97\pm0.03$	&	-	&	[W96] 4771-1038	&	K5	&	14	\\
    WBIS\_053900.5-023939.0$^*$	&	84.7522	&	-2.6608	&	$11.67\pm0.03$	&	-	&	$11.22\pm0.02$	&	-	&	[W96] 4771-1056	&	G3.5	&	11	\\
    WBIS\_053849.2-024125.1$^*$	&	84.7051	&	-2.6903	&	$11.67\pm0.03$	&	-	&	$11.00\pm0.03$	&	-	&	[BNM2013] 93.03 5	&	M6	&	14	\\
    WBIS\_053843.6-023325.4$^*$	&	84.6815	&	-2.5571	&	$11.72\pm0.03$	&	-	&	$10.99\pm0.02$	&	-	&	[W96] rJ053843-0233	&	M1	&	14	\\
    WBIS\_053847.5-023524.9$^*$	&	84.6977	&	-2.5903	&	$11.74\pm0.05$	&	-	&	$10.97\pm0.03$	&	-	&	[BHM2009] SigOri-MAD-34	&	M1	&	14	\\
    WBIS\_053806.7-023022.7$^*$	&	84.5281	&	-2.5063	&	$11.76\pm0.03$	&	-	&	$10.92\pm0.02$	&	-	&	Kiso A-0904 67	&	M1.5	&	14	\\
    WBIS\_053842.3-023714.8$^*$	&	84.6761	&	-2.6208	&	$11.77\pm0.03$	&	-	&	$10.99\pm0.02$	&	-	&	2MASS J05384227-0237147	&	M0	&	14	\\
    WBIS\_053843.0-023614.6$^*$	&	84.6792	&	-2.6040	&	$11.91\pm0.04$	&	-	&	$11.05\pm0.02$	&	-	&	2MASS J05384301-0236145	&	M1	&	11	\\
    WBIS\_053827.3-024509.6$^*$	&	84.6135	&	-2.7527	&	$11.96\pm0.03$	&	-	&	$10.80\pm0.02$	&	-	&	V* V505 Ori	&	M0	&	14	\\
    WBIS\_053834.1-023637.5$^*$	&	84.6419	&	-2.6104	&	$11.98\pm0.03$	&	-	&	$11.33\pm0.02$	&	-	&	[W96] rJ053833-0236	&	M4	&	14	\\
    WBIS\_053845.4-024159.6$^*$	&	84.6890	&	-2.6999	&	$11.99\pm0.03$	&	-	&	$11.33\pm0.03$	&	-	&	V* V595 Ori	&	M1	&	14	\\
    WBIS\_053911.5-023106.5$^*$	&	84.7980	&	-2.5185	&	$11.99\pm0.03$	&	-	&	$11.19\pm0.02$	&	-	&	Haro 5-20	&	M0	&	14	\\
    WBIS\_053833.4-023617.6$^*$	&	84.6390	&	-2.6049	&	$12.05\pm0.03$	&	-	&	$11.30\pm0.02$	&	-	&	2MASS J05383335-0236176	&	M2.5	&	11	\\
    WBIS\_053920.4-022736.8$^*$	&	84.8352	&	-2.4602	&	$12.15\pm0.03$	&	-	&	$11.43\pm0.03$	&	-	&	[BNM2013] 90.02 55	&	M2	&	11	\\
    WBIS\_053831.4-023633.8$^*$	&	84.6309	&	-2.6094	&	$12.17\pm0.03$	&	-	&	$11.47\pm0.02$	&	-	&	Haro 5-11	&	M3.5	&	11	\\
    WBIS\_053827.7-024300.9$^*$	&	84.6156	&	-2.7169	&	$12.19\pm0.03$	&	-	&	$11.45\pm0.02$	&	-	&	[BNM2013] 93.03 46	&	M3	&	11	\\
    WBIS\_053853.2-024352.6$^*$	&	84.7215	&	-2.7313	&	$12.24\pm0.03$	&	-	&	$11.51\pm0.03$	&	-	& 	Mayrit 489165	& 	M1	&	11	\\
    WBIS\_053847.5-022712.0	&	84.6981	&	-2.4533	&	$12.51\pm0.00$	&	$12.43\pm0.00$	&	$12.19\pm0.00$	&	$-0.35\pm0.00$	&	2MASS J05384755-0227120	&	M5	&	5	\\
    WBIS\_053848.3-023641.0	&	84.7012	&	-2.6114	&	$12.51\pm0.00$	&	$12.57\pm0.00$	&	$12.51\pm0.00$	&	$-0.19\pm0.00$	&	2MASS J05384828-0236409	&	M4.5	&	11	\\
    WBIS\_053808.3-023556.2	&	84.5344	&	-2.5990	&	$12.59\pm0.00$	&	$12.45\pm0.00$	&	$12.31\pm0.00$	&	$-0.13\pm0.00$	&	Kiso A-0976 316	&	M2.5	&	11	\\
    WBIS\_053848.7-023616.3	&	84.7028	&	-2.6045	&	$12.62\pm0.00$	&	$12.62\pm0.00$	&	$12.51\pm0.00$	&	$-0.20\pm0.00$	&	[BNM2013] 90.02 90	&	M1.5	&	11	\\
    WBIS\_053925.3-023143.6	&	84.8556	&	-2.5288	&	$12.62\pm0.00$	&	$12.28\pm0.00$	&	$12.24\pm0.00$	&	$0.25\pm0.00$	&	[BNM2013] 90.02 73	&	G3.5	&	11	\\
    WBIS\_053843.3-023200.8	&	84.6805	&	-2.5336	&	$12.65\pm0.00$	&	$12.74\pm0.00$	&	$12.49\pm0.00$	&	$-0.55\pm0.00$	&	[BNM2013] 90.02 74	&	M5	&	11	\\
    WBIS\_053813.2-022608.8	&	84.5550	&	-2.4358	&	$12.70\pm0.00$	&	$12.61\pm0.00$	&	$12.53\pm0.00$	&	$-0.06\pm0.00$	&	2MASS J05381319-0226088	&	M4.5	&	11	\\
    WBIS\_053820.2-023801.6	&	84.5842	&	-2.6338	&	$12.72\pm0.00$	&	$12.49\pm0.00$	&	$12.24\pm0.00$	&	$-0.24\pm0.00$	&	[W96] rJ053820-0237	&	M4/M5	&	4	\\
    WBIS\_053850.4-022647.7	&	84.7099	&	-2.4466	&	$12.72\pm0.00$	&	$12.48\pm0.00$	&	$12.33\pm0.00$	&	$-0.05\pm0.00$	&	[BNM2013] 92.01 11	&	M3.5	&	11	\\
    WBIS\_053903.0-024127.1	&	84.7624	&	-2.6909	&	$12.75\pm0.00$	&	$12.66\pm0.00$	&	$12.66\pm0.00$	&	$0.09\pm0.00$	&	UCAC4 437-010930	&	M2.5	&	11	\\
    WBIS\_053829.1-023602.7	&	84.6213	&	-2.6008	&	$12.79\pm0.00$	&	$12.54\pm0.00$	&	$12.36\pm0.00$	&	$-0.08\pm0.00$	&	[W96] rJ053828-0236	&	M1.5	&	11	\\
    WBIS\_053912.3-023006.4	&	84.8013	&	-2.5018	&	$12.81\pm0.00$	&	$12.69\pm0.00$	&	$12.47\pm0.00$	&	$-0.29\pm0.00$	&	2MASS J05391232-0230064	&	M5	&	11	\\
    WBIS\_053917.2-022543.3	&	84.8215	&	-2.4287	&	$12.81\pm0.00$	&	$12.44\pm0.00$	&	$12.40\pm0.00$	&	$0.30\pm0.00$	&	Kiso A-0904 80	&	M1.5	&	11	\\
    WBIS\_053851.5-023620.6	&	84.7144	&	-2.6057	&	$12.82\pm0.00$	&	$12.92\pm0.00$	&	$12.81\pm0.00$	&	$-0.30\pm0.00$	&	[BNM2013] 90.02 91	&	K5V	&	7	\\
    WBIS\_053847.2-023436.9	&	84.6966	&	-2.5769	&	$12.83\pm0.00$	&	$12.67\pm0.00$	&	$12.46\pm0.00$	&	$-0.22\pm0.00$	&	[W96] pJ053847-0234	&	M4	&	11	\\
    WBIS\_053902.8-022955.8	&	84.7615	&	-2.4988	&	$12.84\pm0.00$	&	$12.69\pm0.00$	&	$12.37\pm0.00$	&	$-0.44\pm0.00$	&	[BNM2013] 90.02 66	&	M4	&	11	\\
    WBIS\_053929.3-022721.0	&	84.8723	&	-2.4558	&	$12.91\pm0.00$	&	$12.62\pm0.00$	&	$12.42\pm0.00$	&	$-0.08\pm0.00$	&	UCAC4 438-011271	&	M0	&	8,11	\\
    WBIS\_053922.9-023333.0	&	84.8453	&	-2.5592	&	$12.91\pm0.00$	&	$12.66\pm0.00$	&	$12.49\pm0.00$	&	$-0.05\pm0.00$	&	[W96] rJ053923-0233	&	M2	&	11	\\
    WBIS\_053827.5-023504.2	&	84.6146	&	-2.5845	&	$12.95\pm0.00$	&	$12.93\pm0.00$	&	$12.64\pm0.00$	&	$-0.50\pm0.00$	&	Kiso A-0976 329	&	M3.5	&	4	\\
    WBIS\_053820.5-023408.9	&	84.5854	&	-2.5691	&	$12.95\pm0.00$	&	$12.97\pm0.00$	&	$12.74\pm0.00$	&	$-0.44\pm0.00$	&	[W96] rJ053820-0234	&	M4	&	11	\\
    WBIS\_053924.4-023401.3	&	84.8515	&	-2.5670	&	$12.98\pm0.00$	&	$12.80\pm0.00$	&	$12.73\pm0.00$	&	$0.05\pm0.00$	&	[BNM2013] 90.02 83	&	M2	&	11	\\
    WBIS\_053907.6-022823.3	&	84.7816	&	-2.4732	&	$12.99\pm0.00$	&	$12.80\pm0.00$	&	$12.52\pm0.00$	&	$-0.33\pm0.00$	&	[W96] rJ053907-0228	&	M3	&	4	\\
    WBIS\_053849.9-024122.8	&	84.7080	&	-2.6897	&	$13.01\pm0.00$	&	$12.98\pm0.00$	&	$12.93\pm0.00$	&	$-0.07\pm0.00$	&	[BNM2013] 93.03 2	&	M3	&	11	\\
    WBIS\_053839.0-024532.1	&	84.6626	&	-2.7589	&	$13.01\pm0.00$	&	$12.78\pm0.00$	&	$12.79\pm0.00$	&	$0.24\pm0.00$	&	2MASS J05383902-0245321	&	M2.5	&	11	\\
    WBIS\_053908.8-023111.5	&	84.7866	&	-2.5199	&	$13.08\pm0.00$	&	$12.78\pm0.00$	&	$12.53\pm0.00$	&	$-0.15\pm0.00$	&	2MASS J05390878-0231115	&	M3	&	11	\\
    WBIS\_053836.9-023643.3	&	84.6536	&	-2.6120	&	$13.13\pm0.00$	&	$13.04\pm0.00$	&	$12.65\pm0.00$	&	$-0.62\pm0.00$	&	2MASS J05383687-0236432	& 	M4.5 	&	8,11	\\
    WBIS\_053841.4-023644.5	&	84.6723	&	-2.6124	&	$13.13\pm0.00$	&	$12.99\pm0.00$	&	$12.62\pm0.00$	&	$-0.56\pm0.00$	&	[BNM2013] 92.01 23	&	M2	&	11	\\
    WBIS\_053859.2-023351.4	&	84.7468	&	-2.5643	&	$13.13\pm0.00$	&	$12.89\pm0.00$	&	$12.65\pm0.00$	&	$-0.21\pm0.00$	&	Haro 5-18	&	M2.5	&	8,11	\\
    WBIS\_053834.6-024108.8	&	84.6442	&	-2.6858	&	$13.14\pm0.00$	&	$12.91\pm0.00$	&	$12.63\pm0.00$	&	$-0.28\pm0.00$	&	[BNM2013] 93.03 29	&	M2	&	8,11	\\
    WBIS\_053817.8-024050.1	&	84.5741	&	-2.6806	&	$13.17\pm0.00$	&	$13.02\pm0.00$	&	$12.85\pm0.00$	&	$-0.16\pm0.00$	&	2MASS J05381778-0240500	&	M5	&	11	\\
    WBIS\_053832.1-023243.1	&	84.6339	&	-2.5453	&	$13.24\pm0.00$	&	$13.39\pm0.00$	&	$12.97\pm0.00$	&	$-0.91\pm0.00$ 	& 	UGCS J053832.13-023243.0	& 	M5	&	11	\\
    WBIS\_053850.8-023626.8	&	84.7116	&	-2.6074	&	$13.25\pm0.00$	&	$13.27\pm0.00$	&	$13.17\pm0.00$	&	$-0.21\pm0.00$	&	[BNM2013] 90.02 362	&	M3	&	8,11	\\
    WBIS\_053904.6-024149.2	&	84.7691	&	-2.6970	&	$13.26\pm0.00$	&	$13.11\pm0.00$	&	$12.99\pm0.00$	&	$-0.07\pm0.00$	&	Mayrit 458140	&	M0	&	8,11	\\
    WBIS\_053914.5-022833.3	&	84.8103	&	-2.4759	&	$13.30\pm0.00$	&	$13.11\pm0.00$	&	$12.92\pm0.00$	&	$-0.15\pm0.00$	&	[BNM2013] 90.02 109	&	M3.5	&	8,4	\\
    WBIS\_053823.5-024131.7	&	84.5981	&	-2.6922	&	$13.31\pm0.00$	&	$13.21\pm0.00$	&	$13.00\pm0.00$	&	$-0.29\pm0.00$	&	2MASS J05382354-0241317	&	M4.5	&	8,11	\\
    WBIS\_053915.8-023650.7	&	84.8159	&	-2.6141	&	$13.31\pm0.00$	&	$13.26\pm0.00$	&	$13.22\pm0.00$	&	$-0.01\pm0.00$	&	[BNM2013] 90.02 146	&	M4	&	8,11	\\
    WBIS\_053846.8-023643.6	&	84.6952	&	-2.6121	&	$13.34\pm0.00$	&	$13.36\pm0.00$	&	$13.22\pm0.00$	&	$-0.29\pm0.00$	&	[BNM2013] 90.02 216	&	M5	&	8,11	\\
    WBIS\_053926.5-022615.4	&	84.8603	&	-2.4376	&	$13.37\pm0.00$	&	$12.96\pm0.00$	&	$12.80\pm0.00$	&	$0.09\pm0.00$	&	2MASS J05392647-0226154	&	M1.5	&	8,11	\\
    WBIS\_053913.5-023739.0	&	84.8061	&	-2.6275	&	$13.44\pm0.00$	&	$13.44\pm0.00$	&	$13.35\pm0.00$	&	$-0.17\pm0.00$	&	2MASS J05391346-0237391	&	M4	&	11	\\
    WBIS\_053823.3-024414.2	&	84.5972	&	-2.7373	&	$13.46\pm0.00$	&	$13.28\pm0.00$	&	$13.10\pm0.00$	&	$-0.16\pm0.00$	&	2MASS J05382332-0244142	&	M4	&	8,11	\\
    WBIS\_053816.1-023804.9	&	84.5671	&	-2.6347	&	$13.47\pm0.00$	&	$13.31\pm0.00$	&	$13.08\pm0.00$	&	$-0.25\pm0.00$	&	2MASS J05381610-0238049	&	M3.5	&	8,11	\\
    WBIS\_053846.0-024523.1	&	84.6916	&	-2.7564	&	$13.51\pm0.00$	&	$13.48\pm0.00$	&	$13.46\pm0.00$	&	$0.00\pm0.00$	&	2MASS J05384597-0245231	&	M5	&	8,11	\\
    WBIS\_053920.2-023825.9	&	84.8343	&	-2.6405	&	$13.57\pm0.00$	&	$13.53\pm0.00$	&	$13.54\pm0.00$	&	$0.07\pm0.00$	&	2MASS J05392023-0238258	&	M5.5	&	11	\\
    WBIS\_053839.7-024019.7	&	84.6655	&	-2.6721	&	$13.66\pm0.00$	&	$13.55\pm0.00$	&	$13.29\pm0.00$	&	$-0.39\pm0.00$	&	2MASS J05383972-0240197	&	M3.5	&	8,11	\\
    WBIS\_053823.1-023649.3	&	84.5962	&	-2.6137	&	$13.69\pm0.00$	&	$13.60\pm0.00$	&	$13.28\pm0.00$	&	$-0.50\pm0.00$	&	2MASS J05382307-0236493	&	M4.5	&	8,14	\\
    WBIS\_053854.9-022858.3	&	84.7289	&	-2.4829	&	$13.73\pm0.00$	&	$13.64\pm0.00$	&	$13.27\pm0.00$	&	$-0.60\pm0.00$	&	2MASS J05385492-0228583	&	M4.5	&	8,9	\\
    WBIS\_053823.3-022534.6	&	84.5972	&	-2.4263	&	$13.74\pm0.00$	&	$13.58\pm0.00$	&	$13.37\pm0.00$	&	$-0.23\pm0.00$	&	V* V2725 Ori	&	M2	&	11	\\
    WBIS\_053826.8-023846.1	&	84.6118	&	-2.6461	&	$14.02\pm0.00$	&	$13.90\pm0.00$	&	$13.56\pm0.00$	&	$-0.52\pm0.00$	&	[BNM2013] 93.03 227	&	M3.5	&	8,11	\\
    WBIS\_053908.1-022844.8	&	84.7837	&	-2.4791	&	$14.04\pm0.00$	&	$13.87\pm0.00$	&	$13.56\pm0.00$	&	$-0.40\pm0.00$	&	2MASS J05390808-0228447	&	M4	&	8,11	\\
    WBIS\_053813.2-022407.5	&	84.5550	&	-2.4021	&	$14.08\pm0.00$	&	$13.97\pm0.00$	&	$13.81\pm0.00$	&	$-0.18\pm0.00$	&	[BZR99] S Ori 13	&	M5.5	&	8,11	\\
    WBIS\_053849.3-022357.5	&	84.7053	&	-2.3993	&	$14.23\pm0.00$	&	$14.11\pm0.00$	&	$13.76\pm0.00$	&	$-0.53\pm0.00$	&	V* V2737 Ori	&	M3.8V	&	8,12	\\
    WBIS\_053848.1-022853.6	&	84.7004	&	-2.4816	&	$14.35\pm0.00$	&	$14.28\pm0.00$	&	$13.86\pm0.00$	&	$-0.72\pm0.00$	&  	[BZR99] S Ori 15	& 	M5.5	&	5,8	\\
    WBIS\_053911.4-023332.8	&	84.7975	&	-2.5591	&	$14.40\pm0.00$	&	$14.30\pm0.00$	&	$13.91\pm0.00$	&	$-0.63\pm0.00$	& 	Mayrit 425070	& 	M5	&	5,8	\\
    WBIS\_053915.1-024047.6	&	84.8129	&	-2.6799	&	$14.45\pm0.00$	&	$14.39\pm0.00$	&	$14.14\pm0.00$	&	$-0.40\pm0.00$	&	[BZR99] S Ori 16	&	M3.5	&	11	\\
    WBIS\_053901.9-023502.8	&	84.7581	&	-2.5841	&	$14.48\pm0.00$	&	$14.07\pm0.00$	&	$13.63\pm0.00$	&	$-0.41\pm0.00$	&	2MASS J05390193-0235029	&	M4	&	8,9	\\
    WBIS\_053838.6-024155.9	&	84.6608	&	-2.6988	&	$14.49\pm0.00$	&	$14.33\pm0.00$	&	$13.96\pm0.00$	&	$-0.51\pm0.00$	&	2MASS J05383858-0241558	&	M5.5	&	6,8	\\
    WBIS\_053825.7-023121.7	&	84.6070	&	-2.5227	&	$14.58\pm0.00$	&	$14.47\pm0.00$	&	$14.10\pm0.00$	&	$-0.57\pm0.00$	&	[BZR99] S Ori 18	&	M3.5	&	8,11	\\
    WBIS\_053835.4-022522.2	&	84.6473	&	-2.4228	&	$14.61\pm0.00$	&	$14.56\pm0.00$	&	$14.15\pm0.00$	&	$-0.72\pm0.00$	&	[BZR99] S Ori 22	& 	M6	&	5,8	\\
    WBIS\_053908.9-023957.9	&	84.7873	&	-2.6661	&	$14.62\pm0.00$	&	$14.68\pm0.00$	&	$14.27\pm0.00$	&	$-0.81\pm0.00$ 	&  	[BZR99] S Ori 25	& 	M6.5/M7.5	&	3,5,8	\\
    WBIS\_053904.5-023835.3	&	84.7687	&	-2.6431	&	$14.68\pm0.00$	&	$14.59\pm0.00$	&	$14.25\pm0.00$	&	$-0.54\pm0.00$	&	[BZR99] S Ori 17	&	M6	&	1,8	\\
    WBIS\_053826.2-024041.3	&	84.6093	&	-2.6781	&	$14.75\pm0.00$	&	$14.69\pm0.00$	&	$14.26\pm0.00$	&	$-0.75\pm0.00$	& 	2MASS J05382623-0240413	& 	M4.5 	&	8,11	\\
    WBIS\_053829.6-022514.2	&	84.6234	&	-2.4206	&	$14.75\pm0.00$	&	$14.75\pm0.00$	&	$14.35\pm0.00$	&	$-0.73\pm0.00$	&  	[BZR99] S Ori 29	& 	M4	&	8,11	\\
    WBIS\_053817.4-024024.2	&	84.5726	&	-2.6734	&	$14.77\pm0.00$	&	$14.69\pm0.00$	&	$14.27\pm0.00$	&	$-0.70\pm0.00$ 	& 	2MASS J05381741-0240242	& 	M5/M7	&	3,8	\\
    WBIS\_053915.8-023826.3	&	84.8157	&	-2.6406	&	$14.89\pm0.00$	&	$14.85\pm0.00$	&	$14.46\pm0.00$	&	$-0.68\pm0.00$	&  	[BZR99] S Ori 26 	& 	M4.5 	&	5	\\
    WBIS\_053825.4-024241.2	&	84.6060	&	-2.7114	&	$15.02\pm0.00$	&	$14.77\pm0.00$	&	$14.37\pm0.00$	&	$-0.48\pm0.00$	&	V* V2728 Ori	&	M7	&	11	\\
    WBIS\_053913.1-023750.9	&	84.8045	&	-2.6308	&	$15.12\pm0.00$	&	$15.09\pm0.00$	&	$14.67\pm0.00$	&	$-0.75\pm0.00$	& 	[BZR99] S Ori 30	& 	M6	&	5,8	\\
    WBIS\_053805.5-023557.1	&	84.5230	&	-2.5992	&	$15.26\pm0.00$	&	$15.16\pm0.00$	&	$14.75\pm0.00$	&	$-0.63\pm0.01$	&	2MASS J05380552-0235571	&	M4.5	&	8,11	\\
    WBIS\_053838.9-022801.7	&	84.6620	&	-2.4671	&	$15.27\pm0.00$	&	$15.08\pm0.00$	&	$14.71\pm0.00$	&	$-0.49\pm0.01$	&	2MASS J05383888-0228016	&	M5V	&	8,11	\\
    WBIS\_053832.4-022957.3	&	84.6352	&	-2.4992	&	$15.33\pm0.00$	&	$15.32\pm0.00$	&	$14.84\pm0.00$	&	$-0.89\pm0.01$	& 	[BZR99] S Ori 39	& 	M6.5	&	3,8	\\
    WBIS\_053923.4-024057.5	&	84.8476	&	-2.6826	&	$16.46\pm0.01$	&	$16.40\pm0.00$	&	$15.92\pm0.01$	&	$-0.83\pm0.02$	&	[BZR99] S Ori 42	& 	M7.5	&	5,8	\\
    WBIS\_053912.9-022453.5	&	84.8037	&	-2.4149	&	$16.71\pm0.01$	&	$16.32\pm0.01$	&	$15.93\pm0.01$	&	$-0.32\pm0.02$	&	2MASS J05391289-0224537	&	M6	&	5,8	\\
    WBIS\_053903.2-023020.0	&	84.7634	&	-2.5056	&	$17.10\pm0.01$	&	$17.17\pm0.01$	&	$16.48\pm0.01$	&	$-1.33\pm0.02$	& 	[BZR99] S Ori 51	& 	M9V	&	2,8,10	\\
    WBIS\_053900.3-023705.9	&	84.7513	&	-2.6183	&	$17.10\pm0.01$	&	$17.08\pm0.01$	&	$16.46\pm0.01$	&	$-1.11\pm0.02$	&  	UGCS J053900.29-023705.7	& 	M8	&	8,10	\\
    WBIS\_053807.1-024321.1	&	84.5297	&	-2.7225	&	$17.13\pm0.01$	&	$16.85\pm0.01$	&	$16.54\pm0.01$	&	$-0.29\pm0.02$	&	[BZR99] S Ori 44	&	M7	&	1	\\
    WBIS\_053814.7-024015.2	&	84.5611	&	-2.6709	&	$17.29\pm0.01$	&	$17.01\pm0.01$	&	$16.43\pm0.01$	&	$-0.78\pm0.02$	& 	[BZR99] S Ori 47	& 	L1	&	2	\\
    WBIS\_053910.8-023714.6	&	84.7950	&	-2.6207	&	$17.47\pm0.01$	&	$17.49\pm0.01$	&	$16.82\pm0.01$	&	$-1.25\pm0.03$ 	& 	[BZR99] S Ori 50	& 	M9	&	2	\\
    WBIS\_053903.6-022536.7	&	84.7650	&	-2.4269	&	$18.37\pm0.03$	&	$18.30\pm0.02$	&	$17.66\pm0.03$	&	$-1.11\pm0.07$	& 	[BZR99] S Ori 58	& 	L0	&	2,8	\\
    WBIS\_053829.5-022937.0	&	84.6230	&	-2.4936	&	$18.76\pm0.03$	&	$18.80\pm0.03$	&	$18.10\pm0.03$	&	$-1.34\pm0.09$	& 	[BMZ2001] S Ori J053829.5-022937	& 	L1/L3.5	&	13	\\
    WBIS\_053857.5-022905.5	&	84.7396	&	-2.4849	&	$18.95\pm0.04$	&	$18.88\pm0.04$	&	$18.18\pm0.04$	&	$-1.24\pm0.12$	& 	UGCS J053857.52-022905.5	& 	L1/L3.5	&	13	\\
    WBIS\_053803.2-022656.7	&	84.5134	&	-2.4490	&	$19.25\pm0.07$	&	$19.09\pm0.05$	&	$18.53\pm0.05$	&	$-0.89\pm0.16$	&	[BMZ2001] S Ori J053803.2-022657	&	L1/L3.5	&	13	\\
    WBIS\_053852.7-022843.7	&	84.7196	&	-2.4788	&	$19.73\pm0.08$	&	$19.38\pm0.05$	&	$19.03\pm0.08$	&	$-0.29\pm0.19$	&	[BZR99] S Ori 61	&	L0	&	2	\\
    WBIS\_053839.2-022805.8	&	84.6631	&	-2.4683	&	$20.03\pm0.09$	&	$20.24\pm0.09$	&	$19.30\pm0.10$	&	$-1.97\pm0.28$	& 	[BZR99] S Ori 68	& 	L5	&	2	\\
    WBIS\_053826.1-022305.0$^{**}$	&	84.6088	&	-2.3847	&	$20.30\pm0.10$ 	&	- 	&	$19.29\pm0.19$ 	&	- 	&	[BZR99] S Ori 65 	&	L3.5 	&	2	\\ 
    WBIS\_053921.0-023033.5	&	84.8374	&	-2.5093	&	$13.28\pm0.00$	&	$13.28\pm0.00$	&	$12.96\pm0.00$	&	$-0.58\pm0.00$	& 	[BZR99] S Ori 3	& 	-	&	8	\\
    WBIS\_053926.8-024258.3	&	84.8615	&	-2.7162	&	$13.32\pm0.00$	&	$13.03\pm0.00$	&	$12.99\pm0.00$	&	$0.19\pm0.00$	& 	2MASS J05392677-0242583	& 	-	&	8	\\
    WBIS\_053844.5-024030.5	&	84.6854	&	-2.6751	&	$13.41\pm0.00$	&	$13.32\pm0.00$	&	$13.00\pm0.00$	&	$-0.52\pm0.00$	& 	2MASS J05384449-0240305	& 	-	&	8	\\
    WBIS\_053905.2-023300.5	&	84.7719	&	-2.5501	&	$13.41\pm0.00$	&	$13.30\pm0.00$	&	$13.01\pm0.00$	&	$-0.43\pm0.00$	& 	2MASS J05390524-0233005	& 	-	&	8	\\
    WBIS\_053901.2-023638.8	&	84.7548	&	-2.6108	&	$13.50\pm0.00$	&	$13.46\pm0.00$	&	$13.36\pm0.00$	&	$-0.15\pm0.00$	& 	2MASS J05390115-0236388	& 	-	&	8	\\
    WBIS\_053850.6-024242.9	&	84.7109	&	-2.7119	&	$13.78\pm0.00$	&	$13.76\pm0.00$	&	$13.67\pm0.00$	&	$-0.14\pm0.00$	& 	2MASS J05385060-0242429	& 	-	&	8	\\
    WBIS\_053848.2-024400.8	&	84.7008	&	-2.7335	&	$13.96\pm0.00$	&	$13.86\pm0.00$	&	$13.74\pm0.00$	&	$-0.12\pm0.00$	& 	2MASS J05384818-0244007	& 	-	&	8	\\
    WBIS\_053833.9-024507.8	&	84.6412	&	-2.7522	&	$14.21\pm0.00$	&	$14.00\pm0.00$	&	$13.69\pm0.00$	&	$-0.36\pm0.00$	& 	2MASS J05383388-0245078	& 	-	&	8	\\
    WBIS\_053925.6-023843.7	&	84.8567	&	-2.6455	&	$15.15\pm0.00$	&	$14.66\pm0.00$	&	$14.31\pm0.00$	&	$-0.16\pm0.01$	& 	HH 446	& 	-	&	8	\\[1ex]
\multicolumn{10}{l}{\textit{Photometric and astrometric candidates-- spectroscopically followed-up in this work $^c$:}}\\[1ex]
     WBIS\_053850.0-023735.5	&	84.7085	&	-2.6265	&	$13.12\pm0.00$	&	$13.17\pm0.00$	&	$13.08\pm0.00$	&	$-0.23\pm0.00$	& 2MASS J05385003-0237354 & Y M5.5 (M6.3$\pm$0.9) & This work \\
      WBIS\_053925.6-023404.2	&	84.8567	&	-2.5678	&	$13.20\pm0.00$	&	$13.09\pm0.00$	&	$12.87\pm0.00$	&	$-0.32\pm0.00$	& 2MASS J05392561-0234042 & Y M5.5 (M6.2$\pm$0.9) & This work\\
      WBIS\_053926.3-022837.7	&	84.8597	&	-2.4771	&	$13.44\pm0.00$	&	$13.25\pm0.00$	&	$13.02\pm0.00$	&	$-0.22\pm0.00$	& 2MASS J05392633-0228376 & Y M5 (M5.6$\pm$0.9) & This work\\
      WBIS\_053911.8-022741.0	&	84.7993	&	-2.4614	&	$13.50\pm0.00$	&	$13.36\pm0.00$	&	$13.12\pm0.00$	&	$-0.30\pm0.00$	& 2MASS J05391183-0227409 & Y M4.5 (M5.3$\pm$0.9) & This work \\
      WBIS\_053845.3-023541.2     &	84.6888	&	-2.5948	&	$13.62\pm0.00$ 	&	$13.63\pm0.00$	&	$13.34\pm0.00$	&	$-0.56\pm0.00$ & Mayrit 21023 &	Y M5.5 (M6.8$\pm$0.9) & This work\\
      WBIS\_053908.2-023228.4 & 84.7842 & -2.5412 & $13.72\pm0.00$ & $13.71\pm0.00$ & $13.34\pm0.00$ & $-0.68\pm0.00$ & [BZR99] S Ori 7 & I M5-M5.5 (M6.4$\pm$0.9) & This work\\
      WBIS\_053835.8-023313.4	&	84.6493	&	-2.5537	&	$13.73\pm0.00$	&	$13.73\pm0.00$	&	$13.32\pm0.00$	&	$-0.74\pm0.00$	& 2MASS J05383582-0233133 & Y M5.5 (M6.9$\pm$0.9) & This work\\
      WBIS\_053853.8-024458.8	&	84.7242	&	-2.7497	&	$15.45\pm0.00$	&	$15.39\pm0.00$	&	$14.90\pm0.00$	&	$-0.84\pm0.01$	& [BMZ2001] S Ori J053853.8-024459 & Y M6-M6.5 (M7.9$\pm$0.9) & This work \\
      WBIS\_053854.9-024033.8	&	84.7288	&	-2.6761	&	$15.67\pm0.00$	&	$15.71\pm0.00$	&	$15.06\pm0.00$	&	$-1.24\pm0.01$ & 2MASS J05385492-0240337 & Y M8 (M8.9$\pm$0.9) & This work	\\
      WBIS\_053812.6-023637.7	&	84.5527	&	-2.6105	&	$15.93\pm0.00$	&	$15.72\pm0.00$	&	$15.23\pm0.00$	&	$-0.71\pm0.01$ & [HHM2007] 395 & Y M6.5-M7 (M7.4$\pm$0.9) & This work	\\
      WBIS\_053913.8-023145.6	&	84.8074	&	-2.5293	&	$17.33\pm0.01$	&	$17.35\pm0.01$	&	$16.77\pm0.01$	&	$-1.09\pm0.03$	& - & Y M8-M8.5 (M9.7$\pm$1.2) & This work \\[1ex]
      WBIS\_053855.4-024120.8$^{d}$	&	84.7309	&	-2.6891	&	$15.19\pm0.00$	&	$15.06\pm0.00$	&	$14.53\pm0.00$	&	$-0.85\pm0.00$ & 2MASS J05385542-0241208 & Y/I M5.5 (M7.1$\pm$0.9) &	This work\\[1ex]
\multicolumn{10}{l}{\textit{Sources identified as members in literature that lack spectral type -- spectroscopically followed-up in this work $^c$:}}\\[1ex]
     WBIS\_053840.5-023327.6	&	84.6689	&	-2.5577	&	$12.97\pm0.00$ &   $13.00\pm0.00$	&	$12.69\pm0.00$	&	$-0.59\pm0.00$ & 2MASS J05384053-0233275 &	Y M5 (M5.5$\pm$0.9) & This work\\
     WBIS\_053843.9-023706.9	&	84.6828	&	-2.6186	&	$13.01\pm0.00$	&	$12.99\pm0.00$	&	$12.82\pm0.00$	&	$-0.31\pm0.00$	& 2MASS J05384386-0237068 & Y M5.5 (M6.6$\pm$0.9) & This work\\
     WBIS\_053851.7-023603.4	&	84.7156	&	-2.6009	&	$13.04\pm0.00$	&	$13.19\pm0.00$	&	$12.98\pm0.00$	&	$-0.54\pm0.00$	& [BNM2013] 90.02 293 & Y M6-M6.5 (M7$\pm$0.9) & This work\\
     WBIS\_053841.6-023028.9	&	84.6733	&	-2.5080	&	$13.05\pm0.00$	&	$13.03\pm0.00$	&	$12.82\pm0.00$	&	$-0.38\pm0.00$	& [BNM2013] 92.01 24 & Y M3-M3.5 ($<$M5) & This work\\
     WBIS\_053849.7-023452.7	&	84.7071	&	-2.5813	&	$13.09\pm0.00$	&	$13.09\pm0.00$	&	$12.83\pm0.00$	&	$-0.49\pm0.00$	 & 2MASS J05384970-0234526 & Y M5 (M6$\pm$0.9) & This work\\
     WBIS\_053845.3-023729.3	&	84.6886	&	-2.6248	&	$13.36\pm0.00$	&	$13.30\pm0.00$	&	$13.14\pm0.00$	&	$-0.25\pm0.00$	& Mayrit 89175 & Y/I M5.5 (M5.9$\pm$0.9) & This work\\
     WBIS\_053847.7-023037.4 & 84.6986  & -2.5104 & $13.44\pm0.00$ & $13.48\pm0.00$ &  $13.11\pm0.00$ & $-0.74\pm0.00$ & [BZR99] S Ori 6 & Y M5.5 (M6.9$\pm$0.9) & This work\\
     WBIS\_053835.3-023313.1 & 84.6470 & -2.5536 & $13.90\pm0.00$ & $13.87\pm0.00$ & $13.49\pm0.00$ & $-0.66\pm0.00$ & [BNM2013] 92.01 46 & Y M5.5 (M7$\pm$0.9) & This work\\
     WBIS\_053841.5-023552.3	&	84.6727	&	-2.5979	&	$14.00\pm0.00$	&	$14.04\pm0.00$	&	$13.65\pm0.00$	&	$-0.75\pm0.00$	& 2MASS J05384146-0235523  & I M6 (M7.7$\pm$0.9) & This work\\
     WBIS\_053842.4-023604.5	&	84.6766	&	-2.6012	&	$14.11\pm0.00$	&	$14.06\pm0.00$	&	$13.69\pm0.00$	&	$-0.63\pm0.00$	& [HHM2007] 683  & I M6 (M6.7$\pm$0.9) & This work \\
     WBIS\_053844.5-024037.7 & 84.6854 & -2.6771 & $14.70\pm0.00$ & $14.75\pm0.00$ & $14.28\pm0.00$ & $-0.93\pm0.00$ & 2MASS J05384448-0240376 & Y M5 (M8.2$\pm$0.9) & This work\\
     WBIS\_053839.8-023220.3 & 84.6657 & -2.5390 & $14.95\pm0.00$ & $15.00\pm0.00$ & $14.41\pm0.00$ & $-1.16\pm0.00$ & 2MASS J05383976-0232203 & Y M6.5-M7 (M9.5$\pm$1.1) & This work\\     
     WBIS\_053818.3-023538.6 & 84.5764 & -2.5940 & $15.23\pm0.00$ & $15.30\pm0.00$ & $14.75\pm0.00$ & $-1.08\pm0.01$ & 2MASS J05381834-0235385 & Y M7 (M8.4$\pm$0.9) & This work\\
     WBIS\_053821.4-023336.3 & 84.5891 & -2.5601 & $15.28\pm0.00$ & $15.22\pm0.00$ & $14.79\pm0.00$ &  $-0.76\pm0.01$ & Mayrit 379292 & I M6 (M8.2$\pm$0.9) & This work\\
     WBIS\_053926.9-023656.1  & 84.8619  & -2.6156 & $15.35\pm0.00$ & $15.28\pm0.00$ & $14.87\pm0.00$ & $-0.70\pm0.01$ & [BZR99] S Ori 36 & Y M6.5 (M8.4$\pm$0.9) & This work\\
     WBIS\_053852.6-023215.5  & 84.7193 & -2.5376 & $16.19\pm0.00$ &  $16.17\pm0.00$ & $15.64\pm0.00$ & $-0.96\pm0.01$ & [MJO2008] J053852.6-023215 & I M6-M6.5 (M8.2$\pm$1.1) & This work\\[1ex]
\multicolumn{10}{l}{\textit{Newly proposed planetary-mass candidates:}}\\[1ex]
     WBIS\_053847.3-023605.6	&	84.6970	&	-2.6015	&	$18.39\pm0.05$	&	$18.63\pm0.03$	&	$17.82\pm0.04$	&	$-1.75\pm0.11$ & -&- &-	\\
     WBIS\_053829.6-022337.8	&	84.6231	&	-2.3938	&	$18.62\pm0.07$	&	$18.54\pm0.05$	&	$17.85\pm0.07$	&	$-1.18\pm0.18$	& -& -&- \\
     WBIS\_053826.8-024514.0	&	84.6118	&	-2.7539	&	$18.68\pm0.07$	&	$18.57\pm0.05$	&	$17.70\pm0.06$	&	$-1.50\pm0.16$ & -&- &	-\\
     WBIS\_053927.4-022759.5	&	84.8642	&	-2.4665	&	$19.80\pm0.13$	&	$19.98\pm0.13$	&	$19.09\pm0.14$	&	$-1.80\pm0.39$	& -& -& -\\
     WBIS\_053918.0-022936.4	&	84.8250	&	-2.4934	&	$20.16\pm0.13$	&	$19.98\pm0.08$	&	$18.98\pm0.10$	&	$-1.67\pm0.28$	& -&- & -\\
\enddata
\tablecomments{CFHT J and H band photometry of sources brighter than J=12.5mag is likely to be saturated so the 2MASS \citep{cutri2003} J and H photometry for these sources are provided. The sources listed as newly proposed planetary-mass candidates have $J>18$mag, beyond the detection capability of IRTF SpeX and require bigger telescopes to confirm their youth and spectral type.\\
$^*$ Sources above the CFHT WIRCam saturation limit whose J and H-band magnitudes are replaced with 2MASS \citep{cutri2003} data.\\
$^{**}$ This object lies at the edge of the CFHT WIRCam survey area and is undetected in our observation. The J and H VISTA photometric data from \citet{pena2012} is provided in the table.\\
$^a$ Names of the sources listed in this column are taken from Simbad.\\
$^b$ Spectral type for these sources are taken from Simbad and their corresponding references are given in the last column.\\
$^c$ Spectral type for these sources were estimated in this work (refer section~\ref{sec:spectral_typing}). The spectral types presented within the parentheses were estimated based on the spectral index following \citet{zhang2018} (refer section~\ref{sec:zj_spt}). All these sources were classified as gravity class VL-G. With the exception of WBIS\_053841.6-023028.9, as the spectral and gravity classification scheme of \citet{zhang2018}
are applicable only for M5 and later spectral type objects. \\
$^d$ This source has been previously identified as a member in \citet{caballero2008}.\\
References: (1) \citet{bejar1999}; (2) \citet{barrado2001}; (3) \citet{bejar2001}; (4) \citet{osorio2002}; (5) \citet{barrado2003}; (6) \citet{caballero2006}; (7) \citet{caballero2007}; (8) \citet{pena2012}; (9) \citet{rigliaco2012}; (10) \citet{canty2013}; (11) \citet{hernandez2014}; (12) \citet{bozhinova2016}; (13) \citet{osorio2017}; (14) \citet{caballero2019}}
\end{deluxetable*}
\end{longrotatetable}

\section{Table of physical parameters}    
\label{sec:app_phypar}
\begin{startlongtable}
\begin{deluxetable*}{cccccccc}
    \tablecaption{Physical parameters of the member sources of $\sigma$ Orionis cluster.}
    \label{tab:physical_parameters}
    \tablehead{
     \colhead{Object ID} & \colhead{RA} & \colhead{Dec} & \colhead{SpT} & \colhead{T$_\mathrm{eff}$} & \colhead{log(L$_\mathrm{bol}$/L$_\mathrm{\sun}$)} & \colhead{Age} & \colhead{Mass}\\
     \colhead{} & \colhead{(deg)} & \colhead{(deg)} & \colhead{} & \colhead{(K)} & \colhead{} & \colhead{(Myr)} & \colhead{(M$_{\sun}$)}\\
     }
    \startdata
     WBIS\_053844.8-023600.2 &	84.6865	&	-2.6000	&	O9.5	&	31900	&	4.72	&	-	&	18.70	\\
    WBIS\_053847.2-023540.6 &	84.6967	&	-2.5946	&	B2	&	20600	&	3.43	&	-	&	7.30	\\
    WBIS\_053845.6-023559.0 &	84.6902	&	-2.5997	&	B2	&	20600	&	3.43	&	-	&	7.30	\\
    WBIS\_053836.5-023312.7 &	84.6523	&	-2.5535	&	B5	&	15700	&	2.77	&	-	&	4.70	\\
    WBIS\_053901.5-023856.4 &	84.7562	&	-2.6490	&	B5	&	15700	&	2.77	&	-	&	4.70	\\
    WBIS\_053834.2-023416.0 &	84.6426	&	-2.5711	&	B7	&	14000	&	2.48	&	-	&	3.92	\\
    WBIS\_053834.8-023415.7 &	84.6450	&	-2.5710	&	A1	&	9300	&	1.49	&	-	&	2.05	\\
    WBIS\_053844.1-023606.3 &	84.6839	&	-2.6018	&	A2	&	8800	&	1.38	&	-	&	1.98	\\
    WBIS\_053827.5-024332.5 &	84.6147	&	-2.7257	&	A2	&	8800	&	1.38	&	-	&	1.98	\\
    WBIS\_053859.5-024508.0 &	84.7481	&	-2.7522	&	F3	&	6660	&	0.37	&	-	&	1.4	\\
    WBIS\_053900.5-023939.0 &	84.7522	&	-2.6608	&	G3.5	&	5680	&	0.05	&	25.12	&	1.10	\\
    WBIS\_053925.3-023143.6	&	84.8556	&	-2.5288	&	G3.5	&	5680	&	-0.33	&	-	&	1.0	\\
    WBIS\_053835.9-024351.1 &	84.6495	&	-2.7309	&	K0	&	5030	&	0.46	&	-	&	0.88	\\
    WBIS\_053918.1-022928.5 &	84.8253	&	-2.4912	&	K0	&	5030	&	0.35	&	-	&	0.88	\\
    WBIS\_053833.7-024414.3 &	84.6405	&	-2.7373	&	K1	&	4920	&	0.57	&	-	&	0.86	\\
    WBIS\_053838.5-023455.0 &	84.6603	&	-2.5820	&	K2	&	4760	&	0.64	&	-	&	0.82	\\
    WBIS\_053807.8-023130.7 &	84.5327	&	-2.5252	&	K3	&	4550	&	0.36	&	1.32	&	1.20	\\
    WBIS\_053853.4-023323.0 &	84.7224	&	-2.5564	&	K3	&	4550	&	0.34	&	1.38	&	1.20	\\
    WBIS\_053844.2-023233.6 &	84.6843	&	-2.5427	&	K4	&	4330	&	0.20	&	1.17	&	1.00	\\
    WBIS\_053835.9-023043.3 &	84.6494	&	-2.5120	&	K4	&	4330	&	0.05	&	2.00	&	1.00	\\
    WBIS\_053835.5-023151.6 &	84.6478	&	-2.5310	&	K4	&	4330	&	0.03	&	2.24	&	1.00	\\
    WBIS\_053932.6-023944.0 &	84.8857	&	-2.6622	&	K5	&	4140	&	0.20	&	0.69	&	0.70	\\
    WBIS\_053838.2-023638.5 &	84.6593	&	-2.6107	&	K5	&	4140	&	0.06	&	1.10	&	0.80	\\
    WBIS\_053834.3-023500.1 &	84.6430	&	-2.5834	&	K5	&	4140	&	0.04	&	1.20	&	0.80	\\
    WBIS\_053907.6-023239.1 &	84.7817	&	-2.5442	&	K5	&	4140	&	0.01	&	1.35	&	0.80	\\
    WBIS\_053844.2-024019.7 &	84.6843	&	-2.6721	&	K5	&	4140	&	-0.02	&	1.48	&	0.80	\\
    WBIS\_053841.3-023722.6 &	84.6720	&	-2.6229	&	K5	&	4140	&	-0.06	&	1.70	&	0.80	\\
    WBIS\_053905.4-023230.3 &	84.7725	&	-2.5417	&	K5	&	4140	&	-0.09	&	2.04	&	0.80	\\
    WBIS\_053911.6-023602.8 &	84.7985	&	-2.6008	&	K5	&	4140	&	-0.12	&	2.24	&	0.80	\\
    WBIS\_053851.5-023620.6	&	84.7144	&	-2.6057	&	K5	&	4140	&	-0.60	&	15.49	&	0.80	\\
    WBIS\_053925.2-023822.0 &	84.8550	&	-2.6394	&	K7	&	3970	&	-0.02	&	0.87	&	0.60	\\
    WBIS\_053849.2-023822.3 &	84.7049	&	-2.6395	&	K7	&	3970	&	-0.05	&	0.98	&	0.60	\\
    WBIS\_053918.8-023053.2 &	84.8285	&	-2.5148	&	K7	&	3970	&	-0.05	&	1.02	&	0.60	\\
    WBIS\_053848.0-022714.2 &	84.7001	&	-2.4540	&	M0	&	3770	&	0.42	&	$<$0.50	&	0.50	\\
    WBIS\_053840.3-023018.5 &	84.6678	&	-2.5051	&	M0	&	3770	&	-0.12	&	0.76	&	0.50	\\
    WBIS\_053831.6-023514.9 &	84.6316	&	-2.5875	&	M0	&	3770	&	-0.12	&	0.76	&	0.50	\\
    WBIS\_053832.8-023539.2 &	84.6368	&	-2.5942	&	M0	&	3770	&	-0.14	&	0.78	&	0.50	\\
    WBIS\_053842.3-023714.8 &	84.6761	&	-2.6208	&	M0	&	3770	&	-0.23	&	1.05	&	0.50	\\
    WBIS\_053827.3-024509.6 &	84.6135	&	-2.7527	&	M0	&	3770	&	-0.30	&	1.41	&	0.50	\\
    WBIS\_053911.5-023106.5 &	84.7980	&	-2.5185	&	M0	&	3770	&	-0.32	&	1.45	&	0.50	\\
    WBIS\_053929.3-022721.0	&	84.8723	&	-2.4558	&	M0	&	3770	&	-0.68	&	6.31	&	0.60	\\
    WBIS\_053904.6-024149.2	&	84.7691	&	-2.6970	&	M0	&	3770	&	-0.82	&	10.72	&	0.60	\\
    WBIS\_053843.6-023325.4 &	84.6815	&	-2.5571	&	M1	&	3630	&	-0.22	&	0.74	&	0.40	\\
    WBIS\_053847.5-023524.9 &	84.6977	&	-2.5903	&	M1	&	3630	&	-0.23	&	0.76	&	0.40	\\
    WBIS\_053843.0-023614.6 &	84.6792	&	-2.6040	&	M1	&	3630	&	-0.30	&	0.95	&	0.40	\\
    WBIS\_053845.4-024159.6 &	84.6890	&	-2.6999	&	M1	&	3630	&	-0.33	&	1.02	&	0.40	\\
    WBIS\_053853.2-024352.6 &	84.7215	&	-2.7313	&	M1	&	3630	&	-0.43	&	1.45	&	0.40	\\
    WBIS\_053806.7-023022.7 &	84.5281	&	-2.5063	&	M1.5	&	3560	&	-0.26	&	0.71	&	0.40	\\
    WBIS\_053848.7-023616.3	&	84.7028	&	-2.6045	&	M1.5	&	3560	&	-0.60	&	2.00	&	0.40	\\
    WBIS\_053829.1-023602.7	&	84.6213	&	-2.6008	&	M1.5	&	3560	&	-0.67	&	2.57	&	0.40	\\
    WBIS\_053917.2-022543.3	&	84.8215	&	-2.4287	&	M1.5	&	3560	&	-0.67	&	2.69	&	0.40	\\
    WBIS\_053926.5-022615.4	&	84.8603	&	-2.4376	&	M1.5	&	3560	&	-0.90	&	6.61	&	0.40	\\
    WBIS\_053920.4-022736.8 &	84.8352	&	-2.4602	&	M2	&	3490	&	-0.42	&	0.98	&	0.30	\\
    WBIS\_053922.9-023333.0	&	84.8453	&	-2.5592	&	M2	&	3490	&	-0.73	&	2.51	&	0.30	\\
    WBIS\_053924.4-023401.3	&	84.8515	&	-2.5670	&	M2	&	3490	&	-0.75	&	2.75	&	0.30	\\
    WBIS\_053841.4-023644.5	&	84.6723	&	-2.6124	&	M2	&	3490	&	-0.81	&	3.47	&	0.40	\\
    WBIS\_053834.6-024108.8	&	84.6442	&	-2.6858	&	M2	&	3490	&	-0.82	&	3.63	&	0.40	\\
    WBIS\_053823.3-022534.6	&	84.5972	&	-2.4263	&	M2	&	3490	&	-1.06	&	9.33	&	0.40	\\
    WBIS\_053833.4-023617.6 &	84.6390	&	-2.6049	&	M2.5	&	3425	&	-0.39	&	0.78	&	0.30	\\
    WBIS\_053808.3-023556.2	&	84.5344	&	-2.5990	&	M2.5	&	3425	&	-0.61	&	1.41	&	0.30	\\
    WBIS\_053903.0-024127.1	&	84.7624	&	-2.6909	&	M2.5	&	3425	&	-0.67	&	1.66	&	0.30	\\
    WBIS\_053839.0-024532.1	&	84.6626	&	-2.7589	&	M2.5	&	3425	&	-0.77	&	2.34	&	0.30	\\
    WBIS\_053859.2-023351.4	&	84.7468	&	-2.5643	&	M2.5	&	3425	&	-0.82	&	2.75	&	0.30	\\
    WBIS\_053827.7-024300.9 &	84.6156	&	-2.7169	&	M3	&	3360	&	-0.45	&	0.78	&	0.30	\\
    WBIS\_053907.6-022823.3	&	84.7816	&	-2.4732	&	M3	&	3360	&	-0.77	&	1.86	&	0.30	\\
    WBIS\_053849.9-024122.8	&	84.7080	&	-2.6897	&	M3	&	3360	&	-0.78	&	1.91	&	0.30	\\
    WBIS\_053908.8-023111.5	&	84.7866	&	-2.5199	&	M3	&	3360	&	-0.81	&	2.14	&	0.30	\\
    WBIS\_053850.8-023626.8	&	84.7116	&	-2.6074	&	M3	&	3360	&	-0.88	&	2.63	&	0.30	\\
    WBIS\_053841.6-023028.9	&	84.6733	&	-2.5080	&	M3	&	3360	&	-0.80	&	2.04	&	0.30	\\
    WBIS\_053831.4-023633.8 &	84.6309	&	-2.6094	&	M3.5	&	3260	&	-0.46	&	0.50	&	0.20	\\
    WBIS\_053850.4-022647.7	&	84.7099	&	-2.4466	&	M3.5	&	3260	&	-0.68	&	1.12	&	0.20	\\
    WBIS\_053827.5-023504.2	&	84.6146	&	-2.5845	&	M3.5	&	3260	&	-0.77	&	1.41	&	0.20	\\
    WBIS\_053914.5-022833.3	&	84.8103	&	-2.4759	&	M3.5	&	3260	&	-0.91	&	2.04	&	0.20	\\
    WBIS\_053816.1-023804.9	&	84.5671	&	-2.6347	&	M3.5	&	3260	&	-0.98	&	2.57	&	0.20	\\
    WBIS\_053839.7-024019.7	&	84.6655	&	-2.6721	&	M3.5	&	3260	&	-1.06	&	3.24	&	0.20	\\
    WBIS\_053826.8-023846.1	&	84.6118	&	-2.6461	&	M3.5	&	3260	&	-1.20	&	5.01	&	0.20	\\
    WBIS\_053915.1-024047.6	&	84.8129	&	-2.6799	&	M3.5	&	3260	&	-1.37	&	8.71	&	0.20	\\
    WBIS\_053825.7-023121.7	&	84.6070	&	-2.5227	&	M3.5	&	3260	&	-1.42	&	10.00	&	0.20	\\
    WBIS\_053849.3-022357.5	&	84.7053	&	-2.3993	&	M3.5	&	3260	&	-1.28	&	6.46	&	0.20	\\
    WBIS\_053834.1-023637.5 &	84.6419	&	-2.6104	&	M4	&	3160	&	-0.40	&	$<$0.50	&	0.17	\\
    WBIS\_053847.2-023436.9	&	84.6966	&	-2.5769	&	M4	&	3160	&	-0.74	&	1.12	&	0.17	\\
    WBIS\_053902.8-022955.8	&	84.7615	&	-2.4988	&	M4	&	3160	&	-0.74	&	1.12	&	0.17	\\
    WBIS\_053820.5-023408.9	&	84.5854	&	-2.5691	&	M4	&	3160	&	-0.79	&	1.26	&	0.17	\\
    WBIS\_053915.8-023650.7	&	84.8159	&	-2.6141	&	M4	&	3160	&	-0.93	&	1.70	&	0.17	\\
    WBIS\_053913.5-023739.0	&	84.8061	&	-2.6275	&	M4	&	3160	&	-0.98	&	1.95	&	0.17	\\
    WBIS\_053823.3-024414.2	&	84.5972	&	-2.7373	&	M4	&	3160	&	-0.99	&	2.00	&	0.17	\\
    WBIS\_053908.1-022844.8	&	84.7837	&	-2.4791	&	M4	&	3160	&	-1.22	&	3.80	&	0.17	\\
    WBIS\_053901.9-023502.8	&	84.7581	&	-2.5841	&	M4	&	3160	&	-1.40	&	5.50	&	0.17	\\
    WBIS\_053829.6-022514.2	&	84.6234	&	-2.4206	&	M4	&	3160	&	-1.51	&	6.76	&	0.15	\\
    WBIS\_053848.3-023641.0	&	84.7012	&	-2.6114	&	M4.5	&	3020	&	-0.63	&	$<$0.50	&	0.15	\\
    WBIS\_053813.2-022608.8	&	84.5550	&	-2.4358	&	M4.5	&	3020	&	-0.71	&	$<$0.50	&	0.15	\\
    WBIS\_053823.5-024131.7	&	84.5981	&	-2.6922	&	M4.5	&	3020	&	-0.95	&	0.93	&	0.13	\\
    WBIS\_053823.1-023649.3	&	84.5962	&	-2.6137	&	M4.5	&	3020	&	-1.10	&	1.70	&	0.13	\\
    WBIS\_053854.9-022858.3	&	84.7289	&	-2.4829	&	M4.5	&	3020	&	-1.12	&	1.78	&	0.13	\\
    WBIS\_053805.5-023557.1	&	84.5230	&	-2.5992	&	M4.5	&	3020	&	-1.73	&	6.31	&	0.09	\\
    WBIS\_053836.9-023643.3	&	84.6536	&	-2.6120	&	M4.5 	&	3020	&	-0.88	&	0.50	&	0.13	\\
    WBIS\_053826.2-024041.3	&	84.6093	&	-2.6781	&	M4.5 	&	3020	&	-1.53	&	3.80	&	0.10	\\
    WBIS\_053915.8-023826.3	&	84.8157	&	-2.6406	&	M4.5 	&	3020	&	-1.58	&	4.47	&	0.10	\\
    WBIS\_053911.8-022741.0	&	84.7993	&	-2.4614	&	M4.5	&	3020	&	-1.03	&	1.45	&	0.13	\\
    WBIS\_053820.2-023801.6	&	84.5842	&	-2.6338	&	M4.5	&	3020	&	-0.71	&	$<$0.50	&	0.15	\\
    WBIS\_053847.5-022712.0	&	84.6981	&	-2.4533	&	M5	&	2880	&	-0.65	&	$<$0.50	&	0.13	\\
    WBIS\_053843.3-023200.8	&	84.6805	&	-2.5336	&	M5	&	2880	&	-0.71	&	$<$0.50	&	0.13	\\
    WBIS\_053912.3-023006.4	&	84.8013	&	-2.5018	&	M5	&	2880	&	-0.77	&	$<$0.50	&	0.13	\\
    WBIS\_053817.8-024050.1	&	84.5741	&	-2.6806	&	M5	&	2880	&	-0.91	&	$<$0.50	&	0.11	\\
    WBIS\_053832.1-023243.1	&	84.6339	&	-2.5453	&	M5	&	2880	&	-0.94	&	$<$0.50	&	0.11	\\
    WBIS\_053846.8-023643.6	&	84.6952	&	-2.6121	&	M5	&	2880	&	-0.98	&	$<$0.50	&	0.10	\\
    WBIS\_053846.0-024523.1	&	84.6916	&	-2.7564	&	M5	&	2880	&	-1.05	&	$<$0.50	&	0.10	\\
    WBIS\_053911.4-023332.8	&	84.7975	&	-2.5591	&	M5	&	2880	&	-1.41	&	0.87	&	0.07	\\
    WBIS\_053838.9-022801.7	&	84.6620	&	-2.4671	&	M5	&	2880	&	-1.75	&	4.47	&	0.06	\\
    WBIS\_053840.5-023327.6	&	84.6689	&	-2.5577	&	M5	&	2880	&	-0.83	&	$<$0.50	&	0.11	\\
    WBIS\_053849.7-023452.7	&	84.7071	&	-2.5813	&	M5	&	2880	&	-0.88	&	$<$0.50	&	0.11	\\
    WBIS\_053926.3-022837.7	&	84.8597	&	-2.4771	&	M5	&	2880	&	-1.02	&	$<$0.50	&	0.10	\\
    WBIS\_053908.2-023228.4	&	84.7842	&	-2.5412	&	M5	&	2880	&	-1.13	&	0.50	&	0.09	\\
    WBIS\_053920.2-023825.9	&	84.8343	&	-2.6405	&	M5.5	&	2920	&	-1.07	&	$<$0.50	&	0.10	\\
    WBIS\_053813.2-022407.5	&	84.5550	&	-2.4021	&	M5.5	&	2920	&	-1.28	&	1.45	&	0.09	\\
    WBIS\_053848.1-022853.6	&	84.7004	&	-2.4816	&	M5.5	&	2920	&	-1.39	&	2.29	&	0.08	\\
    WBIS\_053838.6-024155.9	&	84.6608	&	-2.6988	&	M5.5	&	2920	&	-1.44	&	2.88	&	0.075	\\
    WBIS\_053845.3-023541.2	&	84.6888	&	-2.5948	&   M5.5	&	2920	&	-1.09	&	$<$0.50	&	0.10	\\
    WBIS\_053835.3-023313.1	&	84.6470	&	-2.5536	&	M5.5	&	2920	&	-1.21	&	1.02	&	0.09	\\
    WBIS\_053835.8-023313.4	&	84.6493	&	-2.5537	&	M5.5	&	2920	&	-1.14	&	0.50	&	0.10	\\
    WBIS\_053847.7-023037.4	&	84.6986	&	-2.5104	&	M5.5	&	2920	&	-1.02	&	$<$0.50	&	0.11	\\
    WBIS\_053843.9-023706.9	&	84.6828	&	-2.6186	&	M5.5	&	2920	&	-0.85	&	$<$0.50	&	0.11	\\
    WBIS\_053850.0-023735.5	&	84.7085	&	-2.6265	&	M5.5	&	2920	&	-0.89	&	$<$0.50	&	0.11	\\
    WBIS\_053925.6-023404.2	&	84.8567	&	-2.5678	&	M5.5	&	2920	&	-0.93	&	$<$0.50	&	0.11	\\
    WBIS\_053855.4-024120.8	&	84.7309	&	-2.6891	&	M5.5	&	2920	&	-1.72	&	4.57	&	0.07	\\
    WBIS\_053845.3-023729.3	&	84.6886	&	-2.6248	&	M5.5	&	2920	&	-0.99	&	$<$0.50	&	0.11	\\
    WBIS\_053817.4-024024.2	&	84.5726	&	-2.6734	&	M6	&	2860	&	-1.56	&	1.12	&	0.06	\\
    WBIS\_053849.2-024125.1 &	84.7051	&	-2.6903	&	M6	&	2860	&	-0.32	&	$<$0.50	&	0.13	\\
    WBIS\_053835.4-022522.2	&	84.6473	&	-2.4228	&	M6	&	2860	&	-1.50	&	0.87	&	0.06	\\
    WBIS\_053904.5-023835.3	&	84.7687	&	-2.6431	&	M6	&	2860	&	-1.53	&	0.93	&	0.06	\\
    WBIS\_053913.1-023750.9	&	84.8045	&	-2.6308	&	M6	&	2860	&	-1.70	&	3.63	&	0.06	\\
    WBIS\_053912.9-022453.5	&	84.8037	&	-2.4149	&	M6	&	2860	&	-2.34	&	14.13	&	0.05	\\
    WBIS\_053853.8-024458.8	&	84.7242	&	-2.7497	&	M6	&	2860	&	-1.83	&	4.90	&	0.05	\\
    WBIS\_053851.7-023603.4	&	84.7156	&	-2.6009	&	M6	&	2860	&	-0.87	&	$<$0.50	&	0.11	\\
    WBIS\_053821.4-023336.3	&	84.5891	&	-2.5601	&	M6	&	2860	&	-1.77	&	4.47	&	0.06	\\
    WBIS\_053841.5-023552.3	&	84.6727	&	-2.5979	&	M6	&	2860	&	-1.25	&	0.66	&	0.08	\\
    WBIS\_053842.4-023604.5	&	84.6766	&	-2.6012	&	M6	&	2860	&	-1.30	&	0.89	&	0.075	\\
    WBIS\_053832.4-022957.3	&	84.6352	&	-2.4992	&	M6.5	&	2815	&	-1.79	&	2.19	&	0.05	\\
    WBIS\_053926.9-023656.1	&	84.8619	&	-2.6156	&	M6.5	&	2815	&	-1.80	&	2.82	&	0.05	\\
    WBIS\_053844.5-024037.7	&	84.6854	&	-2.6771	&	M6.5	&	2815	&	-1.54	&	0.74	&	0.05	\\
    WBIS\_053852.6-023215.5	&	84.7193	&	-2.5376	&	M6.5	&	2815	&	-2.14	&	7.41	&	0.04	\\
    WBIS\_053908.9-023957.9	&	84.7873	&	-2.6661	&	M7	&	2683	&	-1.43	&	$<$0.50	&	0.05	\\
    WBIS\_053825.4-024241.2	&	84.6060	&	-2.7114	&	M7	&	2683	&	-1.59	&	$<$0.50	&	0.04	\\
    WBIS\_053807.1-024321.1	&	84.5297	&	-2.7225	&	M7	&	2683	&	-2.43	&	10.00	&	0.03	\\
    WBIS\_053818.3-023538.6	&	84.5764	&	-2.5940	&	M7	&	2683	&	-1.67	&	$<$0.50	&	0.04	\\
    WBIS\_053839.8-023220.3	&	84.6657	&	-2.5390	&	M7	&	2683	&	-1.56	&	$<$0.50	&	0.04	\\
    WBIS\_053812.6-023637.7	&	84.5527	&	-2.6105	&	M7	&	2683	&	-1.95	&	0.78	&	0.03	\\
    WBIS\_053923.4-024057.5	&	84.8476	&	-2.6826	&	M7.5	&	2611	&	-2.20	&	0.95	&	0.02	\\
    WBIS\_053900.3-023705.9	&	84.7513	&	-2.6183	&	M8	&	2539	&	-2.47	&	1.66	&	0.02	\\
    WBIS\_053854.9-024033.8	&	84.7288	&	-2.6760	&	M8	&	2539	&	-1.90	&	$<$0.50	&	0.02	\\
    WBIS\_053913.8-023145.6	&	84.8074	&	-2.5293	&	M8.5	&	2467	&	-2.57	&	1.48	&	0.015	\\
    WBIS\_053910.8-023714.6	&	84.7950	&	-2.6207	&	M9	&	2394	&	-2.61	&	1.02	&	0.01	\\
    WBIS\_053903.2-023020.0	&	84.7634	&	-2.5056	&	M9	&	2394	&	-2.47	&	0.50	&	0.01	\\
    WBIS\_053903.6-022536.7	&	84.7650	&	-2.4269	&	L0	&	2248	&	-2.93	&	1.86	&	0.01	\\
    WBIS\_053852.7-022843.7	&	84.7196	&	-2.4788	&	L0	&	2248	&	-3.47	&	100.00	&	0.04	\\
    WBIS\_053814.7-024015.2	&	84.5611	&	-2.6709	&	L1	&	2102	&	-2.42	&	$<$0.50	&	0.007	\\
    WBIS\_053829.5-022937.0	&	84.6230	&	-2.4936	&	L1	&	2102	&	-3.01	&	1.00	&	0.008	\\
    WBIS\_053857.5-022905.5	&	84.7396	&	-2.4849	&	L1	&	2102	&	-3.08	&	2.00	&	0.009	\\
    WBIS\_053803.2-022656.7	&	84.5134	&	-2.4490	&	L1	&	2102	&	-3.20	&	3.63	&	0.01	\\
    WBIS\_053826.1-022305.0	&	84.6088	&	-2.3847	&	L3.5 	&	1756	&	-3.41	&	1.00	&	0.005	\\
    WBIS\_053839.2-022805.8	&	84.6631	&	-2.4683	&	L5	&	1581	&	-3.20	&	$<$0.50	&	0.004	\\
    WBIS\_053921.0-023033.5	&	84.8374	&	-2.5093	&	-	&	-	&	-	&	-	&	-	\\
    WBIS\_053926.8-024258.3	&	84.8615	&	-2.7162	&	-	&	-	&	-	&	-	&	-	\\
    WBIS\_053844.5-024030.5	&	84.6854	&	-2.6751	&	-	&	-	&	-	&	-	&	-	\\
    WBIS\_053905.2-023300.5	&	84.7719	&	-2.5501	&	-	&	-	&	-	&	-	&	-	\\
    WBIS\_053901.2-023638.8	&	84.7548	&	-2.6108	&	-	&	-	&	-	&	-	&	-	\\
    WBIS\_053850.6-024242.9	&	84.7109	&	-2.7119	&	-	&	-	&	-	&	-	&	-	\\
    WBIS\_053848.2-024400.8	&	84.7008	&	-2.7335	&	-	&	-	&	-	&	-	&	-	\\
    WBIS\_053833.9-024507.8	&	84.6412	&	-2.7522	&	-	&	-	&	-	&	-	&	-	\\
    WBIS\_053925.6-023843.7	&	84.8567	&	-2.6455	&	-	&	-	&	-	&	-	&	-	\\
\enddata
\tablecomments{Refer to section ~\ref{sec:phy_param} for details on each parameter. For the sources spectroscopically observed with IRTF in this work, the best-fitting spectral type estimated in section~\ref{sec:spectral_typing} is given in column 4. The T$_\mathrm{eff}$, L$_\mathrm{bol}$ and mass values for O, B and A spectral type sources are taken from table 5 of \citet{pecaut2013}. For the F to L spectral type sources the T$_\mathrm{eff}$, L$_\mathrm{bol}$, age and mass were estimated as detailed in section~\ref{sec:phy_param}. }
\end{deluxetable*}
\end{startlongtable}


\bibliography{main_final}{}
\bibliographystyle{aasjournal}



\end{document}